\documentclass[utf8]{frontiersFPHY} 

\setcitestyle{square} 
\usepackage{url,hyperref,lineno,microtype,subcaption}
\usepackage[onehalfspacing]{setspace}
\usepackage{bm,graphicx,cancel}
\usepackage{newtxtext,newtxmath,amsmath}
\usepackage{calc}
\usepackage{dcolumn}
\usepackage{mathrsfs}
\usepackage[normalem]{ulem} 
\usepackage{xspace}
\usepackage{xcolor}
\usepackage{bbold}
\usepackage{slashed}
\usepackage{tabu}

\newcommand{\rmd}{\mathrm{d}}
\newcommand{\GeV}{{\rm\ GeV}}

\newcommand{\TeV}{{\rm\ TeV}}
\newcommand{\eg}{{\it e.g.}\xspace}
\newcommand{\ie}{{\it i.e.}\xspace}

\newcommand{\sigmav}{\langle \sigma v \rangle} 
\newcommand{\relic}{ \Omega_{\rm DM} h^2}
\newcommand{\mdm}{m_{\rm DM}}
\newcommand{\mmed}{m_{\rm med}}
\newcommand{\gdm}{g_{\rm DM}}
\newcommand{\gsm}{g_{\rm SM}}
\newcommand{\MET}{{\slashed{E}_T}}
\newcommand{\gglu}{g_{g} }
\newcommand{\ggam}{g_{\gamma} }

\newcommand{\y}{Y_0}

\def\keyFont{\fontsize{8}{11}\helveticabold }
\def\firstAuthorLast{C. Arina} 
\def\Authors{Chiara Arina\,$^{1,*}$}
\def\Address{$^{1}$Centre for Cosmology, Particle Physics and Phenomenology (CP3), Universit\'e catholique de Louvain, Chemin du Cyclotron 2, B-1348 Louvain-la-Neuve, Belgium }
\def\corrAuthor{Chiara Arina}

\def\corrEmail{chiara.arina@uclouvain.be}

\begin{document}
\onecolumn
\firstpage{1}

\title[Constraining dark matter simplified models]{Impact of cosmological and astrophysical constraints on dark matter simplified models} 

\author[\firstAuthorLast ]{\Authors} 
\address{} 
\correspondance{} 
\extraAuth{}

\maketitle

\begin{abstract}
Studies of dark matter models lie at the interface of astrophysics, cosmology, nuclear physics and collider physics. Constraining such models entails the capability to compare their predictions to a wide range of observations. In this review, we present the impact of global constraints to a specific class of models, called dark matter simplified models. These models have been adopted in the context of collider studies to classify the possible signatures due to dark matter production, with a reduced number of free parameters. We classify the models that have been analysed so far and for each of them we review in detail the complementarity of relic density, direct and indirect searches with respect to the LHC searches. We also discuss the capabilities of each type of search to identify regions where individual approaches to dark matter detection are the most relevant to constrain the model parameter space. Finally we provide a critical overview on the validity of the dark matter simplified models and discuss the caveats for the interpretation of the experimental results extracted for these models.

\tiny
 \keyFont{ \section{Keywords:} dark matter theory, particle dark matter, direct searches of dark matter, indirect searches of dark matter, beyond standard model physics, LHC phenomenology} 
\end{abstract}

\section{Introduction}

The presence of dark matter, postulated at the beginning of last century~\citep{Kapteyn1922ApJ,Jeans1922MNRAS,Oort1932BAN,zwicky} (see Refs.~\citep{Bertone:2016nfn,deSwart:2017heh} for a review), has been nowadays confirmed by several observations in cosmology and astrophysics. Besides precision measurements on its abundance from the cosmic microwave background and large scale structures, which state $\Omega_{\rm DM} h^2 = 0.1198 \pm 0.00015$~\citep{Ade:2015xua}, there is only gravitational evidence for this dark component while its nature and properties are completely unknown. Baryons can constitute only the 4\% of the total energy content of the universe, not enough to explain the entire matter content of the universe ($\sim 30\%$). This fact supports a non-baryonic origin for the dark matter particles, most likely arising in models beyond the Standard Model (SM) of particle physics, as SM neutrinos were relativistic in the early universe. Several theoretically motivated extensions of the SM, such as supersymmetry or universal extra-dimensions, provide dark matter candidates which fall into the category of WIMPs (Weakly Interacting Massive Particles). These particles are usually neutral, stable at least on cosmological scale, and with a mass in the GeV - TeV energy range. In this review we will comply with the WIMP paradigm and use WIMPs and dark matter interchangeably, even though other possibilities exist, see {\it e.g.}~\cite{Adhikari:2016bei,Marsh:2015xka} and the references therein.

With the generic hypothesis that WIMPs interact with the SM particles, a multitude of experimental approaches have been undertaken to detect it. These methods range from dark matter searches  in underground detectors~\citep{Akerib:2016vxi,Aprile:2017iyp,Amole:2017dex} via the scattering of WIMPs off nuclei (direct detection), to observations of gamma rays, cosmic rays and neutrinos, produced by dark matter annihilation in astrophysical environments (indirect detection),  see \eg Refs.~\citep{Fermi-LAT:2016uux,PhysRevLett.117.091103,dampe,Aartsen:2016exj}, and dedicated searches for missing energy signals at colliders, see \eg~\citep{Abercrombie:2015wmb,Boveia:2016mrp} (production). Yet, despite the enormous experimental effort, the dark matter detection remains a challenge and our understanding of dark matter properties limited, hence WIMP models can span many orders of magnitude in dark matter masses and interaction strengths. This makes it difficult to efficiently study all possible scenarios and models. It is necessary to find a strategy to combine the maximum amount of available experimental information in the most efficient way to: (i) carve out the dark matter models which are inconsistent with experimental observations; (ii) to highlight the most promising regions for discovery in the model parameter space, in the light of the near future dark matter search program; (iii) to highlight the complementarity among the diverse dark matter search methods. Dark matter simplified models (DMsimps from hereafter) represent a convenient framework where to achieve these objectives, and will be the main focus of the review.

In these past few years, the dark matter program at the LHC has set the trend to follow the avenue of DMsimps~\citep{Abdallah:2015ter,Abercrombie:2015wmb,Boveia:2016mrp,Albert:2017onk}, as compared to the Effective Field Theory (EFT) approach or as compared to the study of complete dark matter models. 

EFT states that the dark matter is the only accessible particle at our experiment, while all the other states that might characterise the dark sector are kinematically unaccessible. EFT is characterised only by the new physics scale $\Lambda$, which makes it easy to compare with direct and indirect detection searches. However the limitations of this approach, at least as far as the LHC searches are concerned~\citep{Goodman:2011jq,Shoemaker:2011vi,MarchRussell:2012hi,Busoni:2013lha,Buchmueller:2013dya,Busoni:2014sya,Busoni:2014haa,Bell:2015sza,DeSimone:2016fbz}, have now been recognised by the theoretical and experimental communities. Basically as soon as the momentum transfer of the process  approaches the value of the new physics scale, EFT breaks down and the micro-physics describing the process needs to be taken into account.  As far as it concerns dark matter direct detection, the momentum transfer is about a few MeV, hence EFT is a well defined framework that can be used unless the mediator mass is of the order the MeV. Dark matter indirect detection lies in between the two cases described above and will be discussed in details in the paper. Notice that nowadays EFT at the LHC is a useful tool to grasp complementary information for instance for high scale~\citep{Belyaev:2016pxe} or for strongly interacting~\citep{Bruggisser:2016nzw} dark matter models.

The opposite approach with respect to EFT stands in considering UV (ultraviolet) complete theoretical models, motivated for instance by solving the hierarchy or the little hierarchy problems, such as supersymmetric models. These models have been and still are being extensively investigated in dedicated study programs, by both the theoretical and experimental communities. The complication arising from such models is the large number of free parameters: at present the dark matter data have not enough constraining power (the only measurement so far being the dark matter relic density) to select specific values of these free parameters of the theory space, hence it is common to end up with degeneracies among the parameters. Conversely, complete models usually feature complex dark sectors with interesting correlations among observables that cannot be reproduced by the EFT or simple models.

These simple models, called DMsimps, are constituted by the addition to the SM particle content of a dark matter candidate, stabilised by imposing a $Z_2$ symmetry, which communicates with at least the SM quarks via one mediator. This minimalistic construction consists in  expanding the EFT interaction by introducing the most important state that mediates the interactions of the dark matter (and of the dark sector) with the SM. Simplified models are typically characterised by three or four free parameters: the dark matter mass $\mdm$, the WIMP-SM $\gdm$ and mediator-SM $\gsm$ couplings (or equivalently the coupling WIMP-SM-mediator $y$) and the mediator mass $\mmed$. So far, they have proven useful to categorise the dark matter searches at the LHC and to set up an easy framework for comparison with direct and indirect searches of dark matter. There are however several caveats emerging from the use of DMsimps in relation with LHC searches and direct/indirect dark matter searches, which are currently driving these models, which might seem purely phenomenological constructions, into more natural bottom-up theoretical models~\citep{Bauer:2016gys}.

The rest of this review is organised as follows. Section~\ref{sec:DMsearch} provides a general overview on the dark matter searches, ranging from cosmology to collider. Section~\ref{sec:results} presents the state of art of current DMsimps, with respect to all the dark matter searches presented in the previous section. A special focus is given to the cosmological and astrophysical constraints, as collider constraints are described in depth in many reviews and recommendation papers (see \eg Refs.~\citep{Boveia:2016mrp,Abdallah:2015ter,Abercrombie:2015wmb,DeSimone:2016fbz,Arcadi:2017kky,Kahlhoefer:2017dnp,Morgante:2018tiq} and the references therein).  In particular Sec.~\ref{sec:schan} considers $s$-channel mediator models and distinguishes the case of spin-0, spin-1 and spin-2 bosons, whereas Sec.~\ref{sec:tchan} reviews the status of $t$-channel models. Section~\ref{sec:caveats} discusses the theoretical caveats of DMsimps, while Sec.~\ref{sec:fut} presents potential avenues for the future. We have tried to present the material in a self-contained form as much as possible, so that the review might serve as an introduction for the beginner and as a reference guide for the practitioner.

\section{Overview on dark matter searches}\label{sec:DMsearch}

\subsection{Cosmological constraints, astrophysical and direct searches}\label{sec:astro}
\begin{figure}[t]
\begin{center}
\includegraphics[width=0.5\textwidth,trim=0mm 0mm 0mm 0mm, clip]{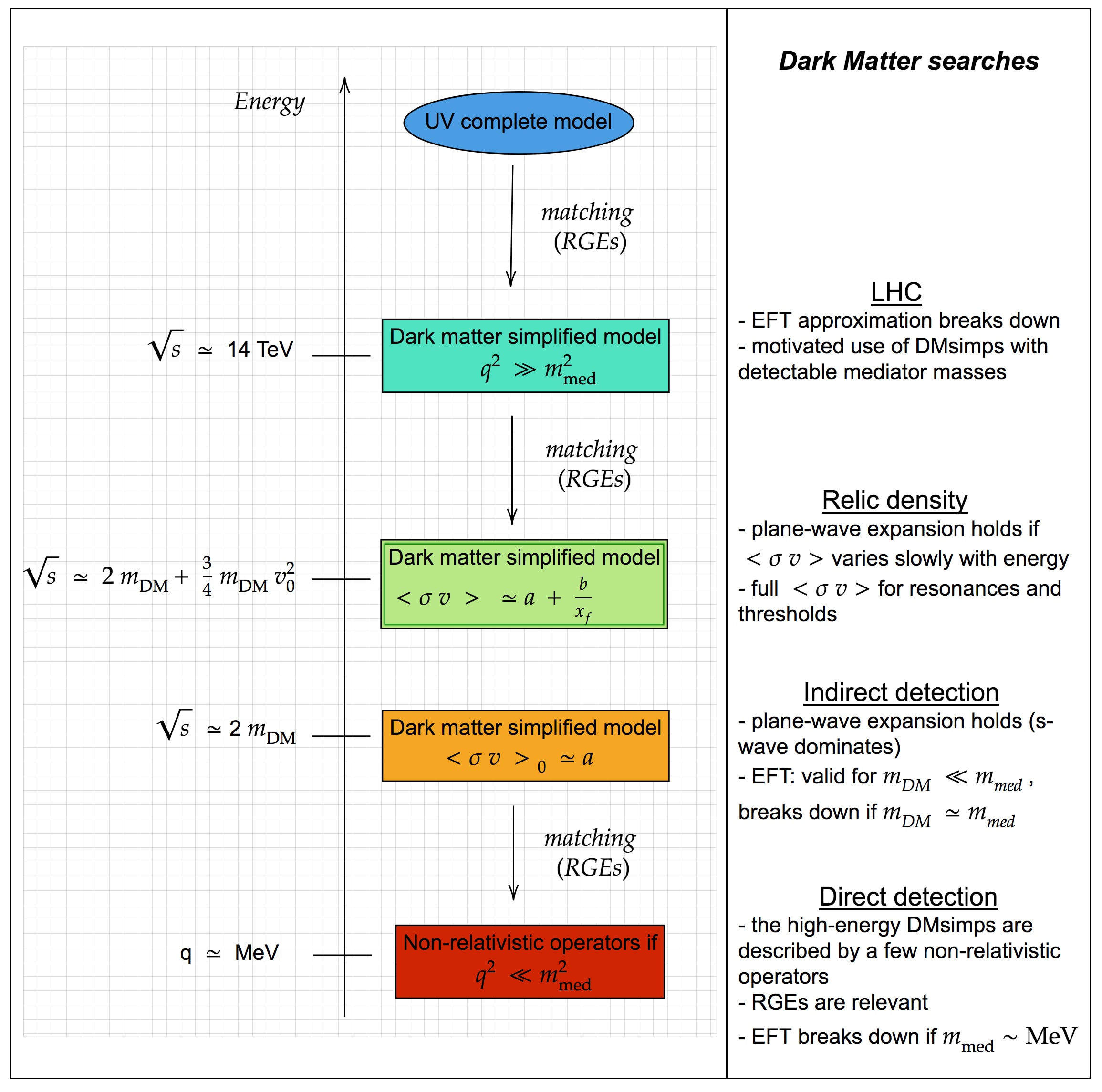}
\end{center}
\caption{Schematic overview of the dark matter searches together with the energy scale typically involved in each of them.}\label{fig:tabE}
\end{figure}

In this section we provide the theoretical basic ingredients to compute cosmological, astrophysical or scattering signals from the DMsimps. For each type of search we discuss whether it is pertinent to use the EFT approximation or if the micro-physics processes should be fully taken into account. A summary plot is provided in Fig.~\ref{fig:tabE}. We also discuss the theoretical assumptions and uncertainties related to each type of search, and how these are interpreted in terms of DMsimps. Finally we briefly review the cosmological constraints on dark matter as well as the several astroparticle searches that are currently running. These constraints will be used to assess the complementarity of searches on the DMsimp parameter space in Sec.~\ref{sec:results}. 

\paragraph*{Dark matter relic density}

In the standard scenario, dark matter is considered a non-relativistic thermal relic in the early universe, which freezes out at $x_f = T/\mdm$ (with $T$ being the temperature of the thermal bath). Its relic abundance  is given roughly by the following approximate solution of the evolution equation (see \eg Refs.~\cite{Srednicki:1988ce,Kolb:1990vq}):
\begin{equation}\label{eq:oh2}
\relic \propto \frac{0.2 \times 10^{-9} \rm GeV}{\sigmav}\,,
\end{equation}
where $\Omega_{\rm DM}$ is as usual the ratio between the dark matter energy density and the critical density of the universe, $h$ is the reduced Hubble parameter ($h = H_0 / 100 \, \rm km\,  s^{-1} \, Mpc^{-1}$, with $H_0$ being the Hubble constant today), and $\sigmav$ is the thermally averaged annihilation cross section. The interaction of the particles needs to be extremely weak in order to achieve $\relic \sim 0.1$:  $\sigmav \sim G_F^2 \mdm^2 \sim 10^{-9} \rm GeV^{-2} \sim 3 \times 10^{-26} \rm cm^3/s$ is just right for GeV-TeV particles\footnote{An upper bound on the WIMP mass of $\mathcal{O}(100) \rm TeV$ stems from the requirement of perturbative unitarity~\citep{Griest:1989wd}, while a lower bound is much more debated and model dependent. In certain models $\mdm > 10 \GeV$ not to spoil recombination~\citep{Ade:2015xua}, for other models $\mdm > 4 \GeV$~\citep{Lee:1977ua}, etc.} to account for all the relic density ($G_F$ is the Fermi constant). This is what is intended with the WIMP paradigm.

The inverse proportionality between $\Omega_{\rm DM}$ and the thermal averaged cross section dictates that: (i) the stronger the interaction rate is, the more depleted is the dark matter number density and as a consequence its relic abundance is too low (`under-abundant' dark matter, namely it contributes to $\relic$ by some \%); (ii) annihilation processes for WIMPs are less efficient, the dark matter particles freeze out at early time and at present time have a significant abundance that matches $\relic$; (iii) the dark matter particles are too feebly interacting, hence they decouple too early and over-close the universe (`over-abundant' dark matter).

If $\sigmav$ varies slowly with energy, it can be expanded in plane waves~\cite{Srednicki:1988ce}:
\begin{equation}\label{eq:exp}
\sigmav  = \langle a + b v^2 + c v^4 + ...\rangle = a + \frac{3}{2} \frac{b'}{x} + \frac{15}{8} \frac{c}{x^2} + ...\,,
\end{equation}
where $b=3/2 b'$. Typically freeze out occurs at $x_f \sim 20-30$ leading to a most probable velocity  $v_0$ of the order of $0.25 c$: corrections proportional to $x^{-1}$ are indeed relevant with respect to the $a$ term and need to be taken into account (notice that the back of the envelop estimate in Eq.~\eqref{eq:oh2} is valid only for a pure $s$-wave $\sigmav$). Additionally, there might be selection rules at play that make the $s$-wave term vanishing. This occurs for several DMsimps, as we will discuss in the next section, which end up having $p$-wave dominated annihilation cross sections.

There are circumstances in which the non-relativistic expansion of $\sigmav$ in Eq.~\eqref{eq:exp} breaks down~\citep{Gondolo:1990dk,Griest:1990kh}:
\begin{itemize}
\item Resonant annihilation: The annihilation cross section is not a smooth function of the centre-of-mass-energy $s$ in the vicinity of an $s$-channel resonance. For $\mdm \leq 2 \mmed$ the additional kinetic energy provided by the thermal bath brings $s$ on top of the resonance and the annihilation cross section increases drastically. Conversely, for $\mdm > 2 \mmed$ the additional kinetic energy brings $s$ even further away from the resonance, hence the annihilation cross section decreases quickly.
\item Opening of new annihilation channels: a fraction of dark matter particles might have a kinetic energy, given by their thermal distribution in the early universe, sufficient to annihilate into heavier particles, which are above the threshold. This again leads to a rapid enhancement of $\sigmav$.
\end{itemize}
In these cases it is necessary to rely on the full computation of the thermally averaged cross section without approximations as well as to solve the complete Boltzmann equation to evaluate precisely $\relic$. This is the standard procedure encoded in the public tools for dark matter (see \eg \texttt{micrOMEGAs}~\citep{Belanger:2018ccd}, \texttt{DarkSUSY}~\citep{Bringmann:2018lay} and\texttt{MadDM}~\citep{Ambrogi:2018jqj}, etc.). As we will see in Sec.~\ref{sec:results}, the model parameter space of DMsimps often features the correct relic density in a tuned-region $\mdm \sim 2 \mmed$, relying on resonant annihilation, and several threshold openings are at play in setting $\relic$. This is schematically resumed in Fig.~\ref{fig:tabE}.

The dark matter relic density is the only precision measurement we have so far. As already anticipated in the introduction, it has been measured with great accuracy by the Planck satellite~\citep{Ade:2015xua}. This measurement, combined with large scale structure data, gives: $\Omega_{\rm DM} h^2 = 0.1198 \pm 0.00015$. The experimental error is at the level of \textperthousand, two orders of magnitude smaller than the associated theoretical error, typically quoted around $\mathcal{O}(10\%)$.

There are a certain number of caveats when considering the relic density as a constraint for DMsimps, which have to be taken into account in the interpretation of the complementarity of searches:
\begin{itemize}
\item DMsimps provide an extension of the SM particle content into the most minimalistic dark sector possible, constituted solely by the dark matter and an extra mediator. If the dark/new physics sector contains more particles, two types of processes can alter the relic density value: (i) there are additional mediators, opening up new annihilation channels including resonance effects; (ii) co-annihilation~\citep{Edsjo:1997bg}, if there are particles heavier but close in mass with the dark matter mass ($\Delta m \lesssim \mathcal(O)(10\%)$). The region allowed by relic density in the DMsimp set up should be considered then as a subset of the whole allowed model parameter space.
\item DMsimps focus particularly on studying and constraining the dark matter-quark couplings, which are accessible at the LHC. However if the dark matter couples to other SM species, additional annihilation diagrams can have a significant impact on the model parameter space allowed by the relic density constraint by opening up new annihilation thresholds. Other couplings, such as dark matter - lepton couplings, start to be considered as well~\citep{Albert:2017onk} in the context of di-lepton searches. In these cases the interpretation of the allowed relic density regions becomes more robust (cfr. the other caveats).
\item The constraint on $\relic$ relies on the assumption that the dark matter is a thermal relic. Even though this hypothesis is attractive as it doesn't depend on the initial conditions in the early universe and set a unique framework to treat SM and dark matter particles, this might not be the case. Other viable assumptions to bring $\relic$ to the observed value, are for instance: (i) the dark matter is non-thermally produced; (ii) the cosmological evolution of our universe is rather different than the one described by the Standard Cosmological model. For example, late-time entropy injection~\citep{Bramante:2017obj} can decrease the dark matter relic density, while late gravitino decays in supersymmetric theories can increase the neutralino relic abundance~\citep{Allahverdi:2012wb}.
\end{itemize}

In Sec.~\ref{sec:results}, we will discuss the combination of dark matter searches in full generality, with and without considering the relic density as relevant constraint. Notice that all caveats described above spoil the model-independent approach of DMsimps, as they rely on the specificity of the dark matter model.

\paragraph*{Dark matter direct detection}

As the dark matter particles move in the Milky Way halo, it is worthwhile to explore the possibility to detect them. This can be done directly in underground terrestrial detectors, sensitive to the nuclear recoil caused by the passing wind of dark matter particles. From a theoretical point of view, in direct detection, the crucial quantity is the scattering cross section of dark matter particles off a nucleon, in a deeply non-relativistic regime. Indeed the momentum transfer in the collision is of the order of a few to tens of MeV, as the speed of the incoming WIMP is of the order of $v \sim 10^{-3} c$. As a consequence, direct detection can be safely treated in term of EFT\footnote{This approximation is satisfied by the DMsimp framework, which typically features mediators heavier than GeV.}, except when the mediator mass connecting the dark matter and the SM quarks becomes of the order of the momentum transfer ($\mmed^2 \sim q^2 \sim \mathcal(O)(10 \rm MeV)$), as resumed in Fig.~\ref{fig:tabE}.

It has been shown that the scattering process between the dark and ordinary matter can be expressed in terms of a limited number of relativistic degrees of freedom, which give rise to a basis of non-relativistic operators. As a matter of fact, any process of elastic scattering between the dark matter and the nucleon can be expressed as a combination of this basis in a unique way, irrespective of the details of the high-energy dark matter model. This basis is constituted by 12 operators, here we report the most relevant for the discussion of Sec.~\ref{sec:results} using the notation of~\citep{DelNobile:2013sia}: 
\begin{eqnarray}
\mathcal{O}^{\rm NR}_1  & = &  \mathbb{1}\,, \qquad \qquad \qquad \mathcal{O}^{\rm NR}_4   =  \mathbf{s}_{\rm DM} \cdot \mathbf{s}_{\rm N} \,,\nonumber \\
\mathcal{O}^{\rm NR}_6  & = & (\mathbf{s}_{\rm DM} \cdot \mathbf{q}) (\mathbf{s}_{\rm N} \cdot \mathbf{q})\,, \qquad \qquad \mathcal{O}^{\rm NR}_8  =  \mathbf{s}_{\rm DM} \cdot \mathbf{v}^{\perp} \,, \qquad \qquad 
\mathcal{O}^{\rm NR}_9   =  i \, \mathbf{s}_{\rm DM} \cdot (\mathbf{s}_{\rm N} \times  \mathbf{q})\,,
\end{eqnarray}
Starting from the DMsimp Lagrangian, which describes the interaction of the dark matter with the quarks, it is necessary first to determine the dark matter-nucleon effective Lagrangian. Secondly, the elastic scattering occurs with the whole nucleus, due to the small WIMP speed in the galactic halo. Therefore, one needs to properly take into account the composite structure of the nucleus which results in the appearance of nuclear form factors in the cross section. Nuclear form factors parametrise the loss of coherence in the scattering with increasing exchanged momentum. In Tab.~\ref{tab:DDeft}, we provide the list of non-relativistic operators relevant for the DMsimps presented in Sec.~\ref{sec:results} and their matching with the matrix element involving the whole nucleus. We refer to Refs.~\citep{Fitzpatrick:2012ix,Cirigliano:2012pq,DelNobile:2013sia,DeSimone:2016fbz} for the rigorous definition of the non-relativistic operator basis and for the detailed direct detection analyses.\footnote{On a side note, except for~\citep{DelNobile:2013sia}, the publicly available dark matter numerical tools do not use the general description of direct detection in terms of non-relativistic operators, at the best of our knowledge at the time of writing.}
\begin{table}[h]
\centering
\caption{\label{tab:DDeft} List of direct detection EFT operators WIMP-nuclei for fermionic and scalar dark matter arising from the DMsimp high-energy interaction Lagrangians discussed in the paper. We provide the matching between these EFT operators and the non-relativistic (NR) operators in the third column. The WIMP-parton coefficients and the transformations from parton level to nuclear EFT operators can be found in \eg~\cite{DelNobile:2013cva}. The dark matter particle is denoted by $X$ if fermionic and by $\Phi$ if scalar, while the nucleus is denoted by $N$ and has a mass $m_N$.  For both Majorana fermions  and real scalars the vector operators vanish, reducing the list of relevant relativistic operators.}
\vspace*{0.5cm}
\tabulinesep=1.mm
\begin{tabu}{|l|c|c|}
\hline
Dark Matter candidate & EFT operator & Matching  \\
\hline
Fermionic & $\bar{X} X \bar{N}N $         &  $4 \mdm m_N \mathcal{O}_1^{\rm NR} $    \\
& $i\,  \bar{X} \gamma_5 X \bar{N}N $   & - $4 m_N \mathcal{O}_{11}^{\rm NR} $          \\
& $i\,  \bar{X}  X \bar{N} \gamma_5 N$  & $4 \mdm  \mathcal{O}_{10}^{\rm NR}$         \\
& $i\,  \bar{X} \gamma_5 X  i\, \bar{N} \gamma_5 N$ & $4  \mathcal{O}_{6}^{\rm NR} $  \\
 & $\bar{X} \gamma^\mu X \bar{N} \gamma_\mu N $  &   $ 8 \mdm (m_N \mathcal{O}_{8}^{\rm NR} + \mathcal{O}_{9}^{\rm NR}) $                     \\
& $i\,  \bar{X} \gamma^\mu \gamma_5 X \bar{N} \gamma_\mu N $  & $ 8 m_N (- \mdm \mathcal{O}_{8}^{\rm NR} + \mathcal{O}_{9}^{\rm NR})$  \\
& $i\,  \bar{X} \gamma^\mu X \bar{N} \gamma_\mu \gamma_5 N$   &  $ -16 \mdm m_N \mathcal{O}_{4}^{\rm NR}  $                                             \\
& $i\,  \bar{X} \gamma^\mu \gamma_5 X  i\, \bar{N} \gamma_\mu \gamma_5 N$   &  $ 32 \mdm m_N \mathcal{O}_{4}^{\rm NR} $                      \\
\hline
Scalar &  $\Phi^\ast \Phi \bar{N}N$  &  $2 \mdm \mathcal{O}_1^{\rm NR}$   \\
 &  i \, $\Phi^\ast \Phi \bar{N} \gamma_5 N$  &  $2  \mathcal{O}_{10}^{\rm NR}$   \\
\hline
\end{tabu}
\end{table}

Concerning the experimental state of art for direct detection, a huge experimental effort has been deployed in the past years, that features nowadays more than 10 different experiments currently running towards unprecedented sensitivities. Several orders of magnitude in the WIMP-nucleus elastic interaction have been constrained by past and current experiments. As far as it concerns spin-independent elastic scattering, which occurs when the dark matter interacts with all the nucleons  (it is proportional to the atomic number of the nucleus, $A^2$), the most notable experiments are XENON1T~\citep{Aprile:2017iyp}, LUX~\citep{Akerib:2016vxi} and PANDAX-II~\citep{Fu:2016ega} for intermediate WIMP masses, CDMSLite~\citep{Agnese:2017jvy} and CRESST-II~\citep{Angloher:2015ewa} at low WIMP masses. XENON1T excludes at 90\% confidence level (CL) WIMP-nucleon cross sections  of about $8 \times 10^{-47} \rm cm^2$ for dark matter masses of 30 GeV. The usual spin-independent scattering cross section corresponds to the operator $\mathcal{O}_1^{\rm NR}$ of Tab~\ref{tab:DDeft}. If present in the underlying particle physics model, this operator dominates over all other non-relativistic operators. Spin-dependent scattering occurs when the dark matter interacts with the spin of the unpaired proton or neutron of the nucleus. PICO 60~\cite{Amole:2017dex} detains the most constraining bound for spin-dependent scattering on proton so far. Only a few experiments are sensitive to the spin-dependent interaction on neutron (mostly dual phase xenon or nobel liquid/gas detectors) and the strongest exclusion bound is held by the LUX~\cite{Akerib:2017kat} experiment. The spin-dependent operator currently considered by the experimental collaborations is $\mathcal{O}_4^{\rm NR}$.  Exclusion limits for the other operators are provided in~\citep{DelNobile:2013sia}, even thought at present these exclusion bounds are a bit outdated. On the experimental side, the XENON collaboration has started to use the non-relativistic operator description and has released exclusion limits based on the XENON100 data~\citep{Aprile:2017aas}.

Direct detection is affected by several astrophysical uncertainties related for instance to the description of the dark matter velocity distribution at the Sun position and to the local dark matter density. There are two different approaches to deal with these uncertainties: either perform a likelihood analysis and marginalise or profile over them, see \eg~\citep{Strigari:2009zb,Arina:2011si,Bertone:2011nj,Arina:2013jma}, either use the so-called halo-independent method, see \eg~\citep{Fox:2010bu,Gondolo:2012rs,DelNobile:2013cva}. In most of the analyses described in Sec.~\ref{sec:results}, astrophysical uncertainties are not taken into account, hence we will not consider this matter any further.

\paragraph*{Dark matter indirect detection}

Dark matter indirect detection relies on the principle that dark matter particles in galactic halos annihilate into SM particles. These SM particles subsequently undergo decays, showering and hadronisation and lead to a continuum flux of cosmic rays, gamma rays and neutrinos. In the case where the dark matter annihilates via loop-induced processes into a pair of photons or a photon and a boson, the signal is characterised by a sharp spectral feature such as a gamma-ray line.  Dark matter annihilation takes place in several astrophysical environments and at different epochs in the evolution of the universe, from cosmological down to solar system scales. As dark matter indirect detection encompasses a large variety of searches, in this review we describe only the searches that have been directly used as complementary probes together with LHC dark matter searches and/or direct detection to constrain DMsimps. Those involve mainly gamma rays, neutrinos and anti-protons at galactic scales. For a detailed review on dark matter indirect detection we refer the reader to \eg~\citep{Cirelli:2015gux,Gaskins:2016cha,Slatyer:2017sev}.

Before going into the details of the specific searches and theoretical predictions, let us mention two generic features concerning the flux of particles produced by dark matter annihilation. This quantity is proportional to
\begin{enumerate}
\item $\sigmav_0$. This is defined as the velocity averaged annihilation cross section computed at present time. Annihilation in galactic halos occurs in a highly non-relativistic regime with an centre-of-mass-energy provided by $\sqrt{s} = 2 \mdm$ as the typical mean velocities characterising the dark matter halo are negligible. For instance in the Milky Way the most probable velocity of dark matter particles is $v_0 \sim 10^{-3} c \sim 230 \, \rm km/s$~\citep{Schoenrich:2009bx}, while it is even lower in dwarf Spheroidal galaxies (dSphs), $v_0 \sim 10^{-5} c \sim 8 \, \rm km/s$~\citep{Bonnivard:2015xpq}, hence in indirect searches the non-relativistic expansion of $\sigmav_0$ in plane waves is a fairly good approximation. The dominant term that is in the reach of current astrophysical probe is the $s$-wave: $\sigmav_0 \simeq a$. If this term is absent due to some selection rule, the model is most likely unconstrained from indirect detection. Notice that the EFT approach remains valid and can be used for $\mdm  \ll \mmed$. This is summarised in Fig.~\ref{fig:tabE}. 
\item ${\rm d} N_f/{\rm d} E_f \equiv \sum_i B_i {\rm d} N^i_f/{\rm d} E_f $. This is defined as the energy spectrum of the particle species $f$ (with $f=\gamma, \nu_l, e^+, \bar{p}$, and $l$ is the neutrino flavour, $l=e,\mu,\tau$) at production where annihilation occurred. The index $i$ runs over all possible annihilation final states of the dark matter model, each of them with a branching ratio $B_i$. The final states are typically SM pairs of particles, however new particles beyond the SM can appear as well, which will subsequently decay into SM particles. We will see in Sec.~\ref{sec:results} that this option is realised in several DMsimps.  

Typically the experimental searches present the limits in a model-independent way, supposing a branching ratio of 100\% into one species of SM particles and assuming that $\relic$ matches the observed value. To compare a specific dark matter model with the experimental exclusion limits, the most rigorous procedure is to recompute the upper bound for that particular model by means of the experimental likelihoods. If this is not possible, one can combine the experimental exclusion bounds after having rescaled them by the appropriate branching ratio. This procedure should be a good approximation provided the energy spectrum of the specific model does not differ too much from the energy spectrum for which each respective upper bound has been computed.  The \texttt{micrOMEGAs} and \texttt{DarkSUSY} numerical tools rely on tabulated energy spectra for all possible SM final states and for dark matter masses ranging from 5 GeV to 100 TeV. The \texttt{MadDM} tool~\citep{Ambrogi:2018jqj} allows to generate the energy spectrum in both model-independent and model-dependent ways for any possible dark matter mass.
\end{enumerate}

Similarly to direct detection, indirect detection is affected by astrophysical uncertainties related to the dark matter density distribution in galactic halos, by the propagation parameters for cosmic rays, etc. Whenever relevant, we will discuss the comparison between different dark matter searches and the indirect detection limits based on different assumptions on the astrophysics.

\subparagraph*{Gamma-ray searches}

The gamma-ray flux from dark matter annihilation from a direction $\psi$ in the sky, averaged over an opening angle $\Delta \psi$, is given by:
\begin{eqnarray}
\frac{{\rmd}\Phi }{\rmd E_\gamma} (E_\gamma, \psi)  =   \frac{\sigmav_0}{2 m_\chi^2}\,  \sum_{i} B_i \frac{{\rmd}N^i_\gamma}{{\rmd}E_\gamma}\,    \frac{1}{4 \pi} \int_\psi \frac{{\rmd} \Omega}{\Delta \psi}\int_\text{los} \rho^2(\psi,l)\,  {\rmd}l \,.
\label{eq:difflux}
\end{eqnarray}
For dark matter particles with distinct particle and antiparticle Eq.~\eqref{eq:difflux} is multiplied by an additional factor of 1/2. The two integrals, over the angle and the line of sight (los), define the astrophysical $J$ factor $\Big( J \equiv \int_\psi {\rmd} \Omega / \Delta \psi \int_\text{los} \rho^2(\psi,l)\,  {\rmd}l \,\Big )$. The $J$ factor encodes the information about the astrophysical environment (experimental window) where annihilation occurs (is sought) and the dark matter density profile.

The most robust gamma-ray constraints at present stem from dSphs, which are dark matter dominated objects~\citep{Mateo:1998wg,Weisz:2011gp,Brown:2012uq,Courteau:2013cjm}. The Fermi-LAT satellite looks for a gamma-ray emission from these Milky Way satellite galaxies, and so far, no excess in gamma rays has been observed.\footnote{There are four dSphs recently discovered by DES~\citep{Abbott:2005bi}, which, taken individually, show a slight excess over the background, of the order of 2$\sigma$. Other analyses, see \eg~\cite{Geringer-Sameth:2015lua,Hooper:2015ula} have pointed out similarly a possible excess over the background. The excess disappears once the data are stacked with the other dSph data.} Hence the Fermi-LAT collaboration has set upper bounds at 95\% CL on the continuum prompt photon flux produced by dark matter annihilation~\citep{Fermi-LAT:2016uux,Ackermann:2015zua}. From these bounds, it has publicly released upper limits for the annihilation rate into $b\bar{b}$ and $\tau^+\tau^-$ final states as a function of the dark matter mass. The $b\bar{b}$ channel is an example of `soft' channel that produces photons mostly from the decay of neutral pions produced in hadronisation, while the $\tau^+\tau^-$ is a `hard' channel that generates photons from final state radiation, scaling as $E^{-1}$, on top of the photons coming from $\pi^0$ decays. The Fermi-LAT team has performed a stacked likelihood analysis for 45 dSphs with high-confidence evaluation of the $J$ factors (to reduce the astrophysical uncertainties). The resulting profile  function for each dSph has been released publicly and can be used to compare for instance DMsimps with dSphs data from the 6 years Fermi-LAT data (Pass 8 event reconstruction algorithm)~\citep{fermilike}. These likelihood functions have been implemented in the last \texttt{MadDM} version, see~\citep{Ambrogi:2018jqj} for details, and can be used for any generic dark matter model. 
In addition to the prompt photon flux, there are also contribution from inverse Compton scattering or synchrotron radiation generated by charged propagating particles. These are often neglected while computing the exclusion limits on the dark matter annihilation rate, however could have an impact for $\mdm \geq 100 \GeV$. Hence the exclusion bounds for large dark matter masses should be regarded as conservative.

Another search, used in the complementarity framework of DMsimps, looks for gamma-ray spectral features towards the Galactic Centre. These spectral features encompass gamma-ray lines, narrow boxes (see \eg~\citep{Ibarra:2015tya}) and sharp edges in the prompt photon energy spectrum coming for instance from internal bremsstrahlung processes (see \eg~\citep{Giacchino:2013bta,Toma:2013bka}). The most constraining exclusion limits on the dark matter annihilation rate into gamma-ray lines are provided by the Fermi-LAT satellite~\citep{Ackermann:2015lka} for $\mdm < 500 \GeV$ and the HESS telescope for dark matter masses up to 25 TeV~\citep{Abdalla:2016olq,Abramowski:2013ax}. These searches suffer of large astrophysical uncertainties related to the dark matter density profile, included in the $J$ factor, and to the background modeling of the Galactic Centre.~\footnote{In this review we do not consider the Galactic Center excess at low dark matter masses. For details, we refer the reader to \eg ~\citep{Gaskins:2016cha} and the references therein.}

\subparagraph*{Neutrino searches}

If dark matter particles scatter in heavy astrophysical bodies such as the Sun, they can lose enough energy to become gravitationally trapped inside it. Dark matter particles start to accumulate in the center of these celestial bodies, where subsequently dark matter annihilation sets in (see \eg~\citep{Steigman:1978vs,Silk:1985ax,Press:1985ug,Gould:1987ir,Ritz:1987mh,Kamionkowski:1991nj,Jungman:1995df,Bergstrom:1996kp,Gondolo:2004sc,Blennow:2007tw,Peter:2009mk,Sivertsson:2012qj}). In the Sun, constituted primarily by hydrogen, the dark matter capture occurs mainly by spin-dependent elastic scattering (even thought the spin-independent scattering on nucleons, $\sigma^{\rm SI}_n$, can also play a role, as it is enhanced by the $A^2$ term for heavy nuclei~\citep{Gondolo:2004sc}). The Sun is opaque to all dark matter annihilation products but neutrinos, which can escape the Sun surface and be detected by Earth based telescopes such as IceCube and Super-Kamiokande~\citep{Choi:2015ara}. The annihilation rate can become large enough to lead to an equilibrium between dark matter capture and annihilation. In this case $\sigmav_0$ and the elastic cross section on proton, $\sigma^{\rm SD}_p$, become two related quantities that can be trade one for the other. This assumption is used for computing experimentally the exclusion bounds on the WIMP-nucleon elastic cross section. The IceCube collaboration has set stringent upper limits, competitive with those of direct detection searches for spin-dependent scattering~\citep{Aartsen:2012kia,Aartsen:2016exj}, by the non observation of GeV-TeV scale neutrinos coming from the Sun direction. The exclusion bounds publicly released, at 90\% CL, are based on IceCube data with 79 strings including DeepCore and are given for the following final states, `hard' channels ($W^+W^-$, $\tau^+\tau^-$, $ZZ$, $\nu\bar{\nu}$) and `soft' channels ($b\bar{b}$, $t\bar{t}$, $gg$ and $hh$). 

The equilibrium assumption helps in the interpretation and comparison of dark matter exclusion limits coming from direct and indirect detection in terms of WIMP-quark coupling; this is particularly appreciable for DMsimp models, which often do include only these couplings. There is however an emergent caveat: direct detection experiments have pushed the upper bound on the spin-independent and spin-dependent cross-section to lower and lower values for which the equilibrium assumption starts to break down~\citep{Arina:2017sng}. Depending then on the size of $\sigmav_0$ and $\sigma^{\rm SI, SD}_n$, the useful representation of exclusion bounds in terms of elastic scattering might not provide anymore a correct physical interpretation.

\subparagraph*{Anti-proton searches}

Searches for dark matter annihilation products in local charged cosmic-ray fluxes can be highly sensitive, especially due to the low backgrounds for antimatter produced by astrophysical processes. A major challenge for these searches is the identification of the locations of the sources of cosmic rays due to their propagation throughout the Milky Way, conversely to the case of gamma rays and neutrinos, which do not diffuse and trace their source. Anti-protons have been recognised as important messengers not only for cosmic ray physics but constitute one of the primary channels in the dark matter searches~\citep{Silk:1984zy,Bertone:2010zza}. This idea has been further reinforced by the data released recently by the AMS 02 satellite~\citep{PhysRevLett.117.091103}, which have an amazing statistical precision and extend up to 450 GeV. The authors of~\citep{Giesen:2015ufa} and~\citep{Cuoco:2017iax} have provided an analysis of these data in terms of exclusion limits for the dark matter velocity averaged annihilation cross section as a function of $\mdm$ at 95\% CL for the $b\bar{b}$, $gg$, $q\bar{q}$, $t\bar{t}$, $\mu^+\mu^-$, $W^+W^-$, $hh$ and $\gamma\gamma$ final states. These bounds (especially $b\bar{b}$) are used to assess the constraining power of anti-proton searches for DMsimps in some of the analyses presented in Sec.~\ref{sec:results}.

The exclusion limits on the dark matter annihilation rate from anti-protons suffer of very large astrophysical uncertainties. The exclusion limits can fluctuate upwards or downwards by one order of magnitude at low dark matter masses, mainly because of uncertainties in the propagation parameters in our galaxy and of solar modulation. The choice of the dark matter density profile is not the main cause of the lack of precision. For details we refer to~\citep{Giesen:2015ufa,Cirelli:2015gux,Cuoco:2017iax} and the references therein.

\subsection{LHC dark matter searches}\label{sec:lhc}
In this section we summarise very briefly the main dark matter searches pursued by the LHC experimental collaborations. For a detailed information, we refer the reader to \eg~\citep{Abercrombie:2015wmb,DeSimone:2016fbz,Albert:2017onk,Kahlhoefer:2017dnp,Morgante:2018tiq} and the references therein.

During the LHC Run 2, ATLAS and CMS have gone the avenue of dark matter simplified models to classify and categorise all possible final states arising in the dark matter search program. This method has been validated by the Dark Matter forum~\citep{Abercrombie:2015wmb} and further supported by the LHC Dark Matter Working Group, established as the successor of the Dark Matter Forum.\footnote{We chose not to provide any reference here for the specific searches conducted by ATLAS and CMS, and to provide the references in the next section, referring to the data sets actually used in the analyses discussed in this review.} 

The main bulk of dark matter searches at colliders is constituted by signatures with missing transverse energy ($\MET$) in the final state, due to the pair-produced dark matter particles which elusively leave the detector with no trace. Namely the mediator, produced by Drell-Yan or gluon fusion, decays invisibly into a pair of dark matter particles. The apparent imbalance of energy is compensated by an energetic jet, photon, etc, coming from typically initial state radiation. These are the most relevant searches for DMsimps undertook so far by the ATLAS and CMS collaborations and are called mono-$X$ + $\MET$ searches, where $X$ stems for a jet, a photon, a vector boson, a Higgs, and multi-jets (from 2 to 6 jets) + $\MET$. All these searches require $2 \mdm << \mmed$ and possible that the mediator has a large branching ratio into dark matter and SM particles (large $\gdm$ and especially large $\gsm$). Once these conditions are met, the searches are not very sensitive to the actual mass of the dark matter particle. This is the reason why  LHC searches are more sensitive to very light dark matter masses, close the $\mathcal{O}(1)$ GeV with respect to direct detection searches~\citep{Boveia:2016mrp}. Additionally to mono-X + $\MET$ searches, a certain number of DMsimps can be constrained by recasting searches in supersymmetric simplified models or by $t \bar{t} + \MET$ searches.

Both the experimental and theoretical communities have recognised that resonance searches for the mediator can be as powerful as the $\MET$ signals in DMsimps, or in some case be even more constraining, see \eg~\citep{Arina:2016cqj,Albert:2017onk}. These searches are based on the principle that, after its production by proton collisions, the mediator does not necessarily  decay into dark matter particles but can decay back into SM final states. This is always the case for $\mmed < 2 \mdm$, as the invisible decay channel is closed; it is also satisfied for $\gsm > \gdm$, condition that leads to a small branching ratio into dark matter particles and a large branching ratio into visible SM species. Besides the two requirements above these searches as well are not very sensitive to the dark matter exact mass value.  In general the most relevant resonance searches, depending on the specific of the DMsimp, are $t\bar{t}$, $4$ tops, di-photons, di-leptons and di-jets. The sensitivity of each search depends on the specificity of the DMsimp under investigation. For instance, di-jet signals are irrelevant for scalar mediators, while $t\bar{t}$ pair production and di-photons reveal very useful~\citep{Arina:2016cqj}. Conversely spin-1 mediators are easily probed via di-jets and mono-X signatures~\citep{Chala:2015ama,Pree:2016hwc}.

Notice that the discovery of an anomalous signals in a mono-X + $\MET$ search at the LHC would not imply the discovery of dark matter, contrary to the case of direct and indirect detection searches. Hence a potential discovery at colliders needs to be supported by further evidence in direct or indirect searches, in order to fully identify the dark matter candidate. On the other hand, in case of new findings, LHC is  able to provide an accurate characterisation of the new mediator particle, while direct and indirect detection are more loosely sensitive to it.

\section{Current status of dark matter simplified models}\label{sec:results}

Since the start of the LHC Run 2 and the publication of the DM forum recommendations~\citep{Abercrombie:2015wmb}, the number of works studying DMsimps has increased exponentially. DMsimps have been adopted for their minimalistic structure to provide the SM with a dark matter particle, in the sense that they represent the minimal extension of the EFT approach used in the LHC Run 1 dark matter searches. The EFT operators are opened up by introducing a particle mediating the interaction between the dark matter and the SM particles (the so-called mediator). They are simple enough to allow the LHC experimental collaborations to categorise all possible dark matter signals they can give rise to.  A general classification stems from the class of vertices that characterise the model: Lagrangians featuring WIMP-WIMP-mediator and SM-SM-mediator type interactions identify models with an $s$-channel mediator, while  Lagrangians characterised by WIMP-SM-mediator interactions define a $t$-channel mediator.   In $s$-channel models, the mediator is always a colour singlet, while in $t$-channel models it can be either a coloured particle or a colour singlet (even though this second possibility, is less appealing for the collider phenomenology). Nonetheless, the definition of DMsimp is not unique, especially as far as it concerns the mediator nature. Some works consider Higgs portal models as part of the DMsimp category (see \eg~\citep{Abdallah:2015ter,DeSimone:2016fbz}), while others do not include the SM Higgs boson in this context~\citep{Abercrombie:2015wmb,Boveia:2016mrp}. For the rest of the section we will use the definition of DMsimp as provided in~\citep{Abercrombie:2015wmb,Boveia:2016mrp}:
\begin{itemize}
\item There can be only one new mediator at a time that defines the interaction between the dark matter and the SM quarks. Namely the dark matter and the mediator are the only particle accessible by current experiments. The presence of additional new particles in the dark sector is assumed not to modify sensibly the physics described by DMsimps. This assumption allows to introduce a very limited set of new free parameters (typically four). The mediator can have spin-0, spin-1/2, spin-1 and spin-2. The category of scalar mediators, however, does not include the Higgs boson (and no mixing with it is considered). We will comment on Higgs portal models in Sec.~\ref{sec:caveats}.
\item The new interaction should not violate the exact and approximate accidental global symmetries of the SM.  For instance this means that baryon and lepton number conservation of the SM should be preserved by this interaction. Additionally, the new mediating particle can produce large flavour violating effects. By enforcing that the flavour structure of the couplings between the dark matter and the ordinary particles follow the same structure as in the SM, it is ensured that DMsimps do not violate flavour constraints. This assumption is called Minimal Flavour Violation (MFV)~\citep{D'Ambrosio:2002ex}, for a detailed discussion see \eg~\citep{Abdallah:2015ter}.
\item Another recommendation concerns the nature of the dark matter particle. It is suggested to consider Dirac fermionic candidates only, because LHC searches are rather insensitive to the spin of the dark matter particles. As the $\MET$ searches are based on cut-and-count analyses, minor changes in the kinematic distributions of the visible particle are expected to have little effect on these analyses, besides the fact the Majorana particles forbid some processes allowed for Dirac particles. However, whenever possible, we will review cases that go beyond the Dirac fermionic dark matter assumption, as the dark matter annihilation and elastic scattering cross sections do depend on the dark matter spin. Different selection rules are at play depending whether the dark matter is a real scalar, a complex scalar, a Dirac or Majorana fermion, leading to suppressions or enhancements of direct or indirect detection signals. These selection rules change drastically the complementarity picture of dark matter searches and need to be considered and investigated further. Table~\ref{tab:Dmnat} provides a summary of the sensitivity of each dark matter search as a function of the DMsimp and of the spin of the dark matter particle, considered in this review.
\end{itemize}
\begin{table}
\centering
\caption{\label{tab:Dmnat}Schematic summary of the complementarity of dark matter searches for the DMsimps, taking into account the spins and nature of both mediator and dark matter particles. In the table, S = scalar, P = pseudo-scalar, V = vector, A = axial-vector, F = fermion, D = Dirac, M = Majorana, DM = dark matter, $Y$ = mediator, DD = direct detection, SI = spin-independent, SD = spin-dependent, ID = indirect detection. OK means that the corresponding signal is in the reach of current and near future experiments, while NO means that the predictions are far below the experimental sensitivities, and NA means that there are no actual studies to assess the experimental reach, to the best of our knowledge. The analytic expressions for the annihilation and scattering cross sections can be found \eg~\citep{DeSimone:2016fbz,Albert:2017onk,Lee:2014caa}. For each DMsimp, the minimal model is considered, which entitles only couplings between the mediator and the SM quarks, as described in Sec.~\ref{sec:results} of this review. The only exception is the spin-2 model, where the mediator communicates with all SM fields.}
\vspace*{0.5cm}
\tabulinesep=1.mm
\begin{tabu}{|c | c | c | c | c | c | c | c |}
\hline
$Y$ Spin & DM spin & \multicolumn{2}{c|}{DD} & \multicolumn{2}{c|}{ ID $\sigmav_0$} & \multicolumn{2}{c|}{LHC searches} \\
\hline
& & SI & SD & $s$-channel  &   $t$-channel  & $\MET$  & Resonance  \\
\hline
S & S & OK & NO & helicity suppressed & $s$-wave & large $\gdm$, $\gsm$  &  OK \\
 &  &  &  & $\propto m_f^2$ &  &  &   \\
 & D & OK & NO & $p$-wave & $p$-wave & large $\gdm$, $\gsm$ & OK \\
 \hline
P & D & NO & NO & $s$-wave & $p$-wave &  large $\gdm$, $\gsm$ & OK \\
 \hline
S-P & D & OK & NO & $p$-wave & $s$-wave & large $\gdm$, $\gsm$ &  OK \\
 &  & (if $\gsm$ is large)  &  &  & & &   \\
 \hline
 V & S & OK & NO & $p$-wave & / & OK &  OK \\
 & D & OK & NO & $s$-wave & $p$-wave & OK & OK \\
 \hline
 A & D & OK & OK & helicity suppressed  & $s$-wave &  OK &  OK \\
 & M & NO & OK & helicity suppressed   & $s$-wave &  OK & OK \\
 \hline
2  & S & NA & NO & $s$-wave & $s$-wave & large  $\gdm$, $\gsm$   &  OK \\
 & F & NA & NA & $p$-wave & $s$-wave &   large   $\gdm$, $\gsm$  & OK \\ 
 & V & NA & NA & $s$-wave & $s$-wave &  large    $\gdm$, $\gsm$  & OK \\  
\hline
\end{tabu}
\end{table}

Most of DMsimps considered in this review have been implemented in \texttt{FeynRules}~\citep{Alloul:2013bka} and are publicly available for download in the repository of the \texttt{DMsimp} framework~\citep{dmsimp}. DMsimps for $s$-channel mediators include three different choices for the spin of the WIMP (Dirac fermion, real scalar and complex scalar for spin-0 and spin-1 mediators, and real scalar, Dirac fermion and vector dark matter for spin-2 mediators). Typically, the numerical tools used to compute the dark matter relic density and  astrophysical constraints are \texttt{micrOMEGAs}~\citep{Belanger:2018ccd} and \texttt{MadDM}~\citep{Ambrogi:2018jqj}. In the \texttt{MadGraph\_aMC@NLO} platform~\citep{Alwall:2011uj,Alwall:2014hca}, one-loop and NLO (next-to-leading order) computations in QCD and EW interactions can be automatically performed in models beyond the SM. This framework allows to compute accurate and precise predictions for production cross sections and distributions of dark matter particles produced at the LHC in association for instance with a mono-jet, mono-photon, mono-Z or mono-Higgs (see \eg~\citep{Backovic:2015soa,Mattelaer:2015haa,Arina:2016cqj,Das:2016pbk}). It is known that higher order QCD corrections impact not only the production rate but also the shape of the distributions. Most of $s$-channel DMsimps do include NLO corrections to the matrix elements and parton shower matching and merging. Indeed these higher order terms pertain only to the initial state and originate only from SM processes, hence they can be factorised with respect to the leading order (LO) process accounting for the production of the uncoloured mediator and dark matter particles. Conversely the implementation of NLO corrections into $t$-channel DMsimp is much more involved, due to the coloured nature of the mediator, which do not allow anymore to factorise initial and final state corrections. Typically $t$-channel DMsimps are LO models, unless stated otherwise. The NLO DMsimps (implemented with \texttt{NloCT}~\citep{Degrande:2014vpa}) are also available at the \texttt{DMsimp} framework webpage~\citep{dmsimp}.  As far as it concerns the DMsimp predictions for relic density, direct and indirect detection, NLO corrections are typically not considered. The automatisation of loop-induced, one-loop and NLO processes is currently under development in a future release of \texttt{MadDM}, which is now a \texttt{MadGraph\_aMC@NLO} plugin and hence inherits all its features, including the capabilities of automatically generate the above-mentioned processes for dark matter observables.

As the literature about DMsimps is vast, we consider and discuss only a few selected representative papers, whereas we try to be as exhaustive as possible with the references. In the following sections we provide the interaction Lagrangian for DMsimps we consider and the relevant details for the analyses we review. We take into consideration in general only mediator-quark couplings; couplings to leptons or other SM particles are switched on whenever relevant.

\subsection{$s$-channel mediator models}\label{sec:schan}
\subsubsection{Spin-0 mediator}\label{sec:spin0}
\begin{figure}[t]
\begin{center}
\includegraphics[width=0.7\textwidth,trim=0mm 0mm 0mm 0mm, clip]{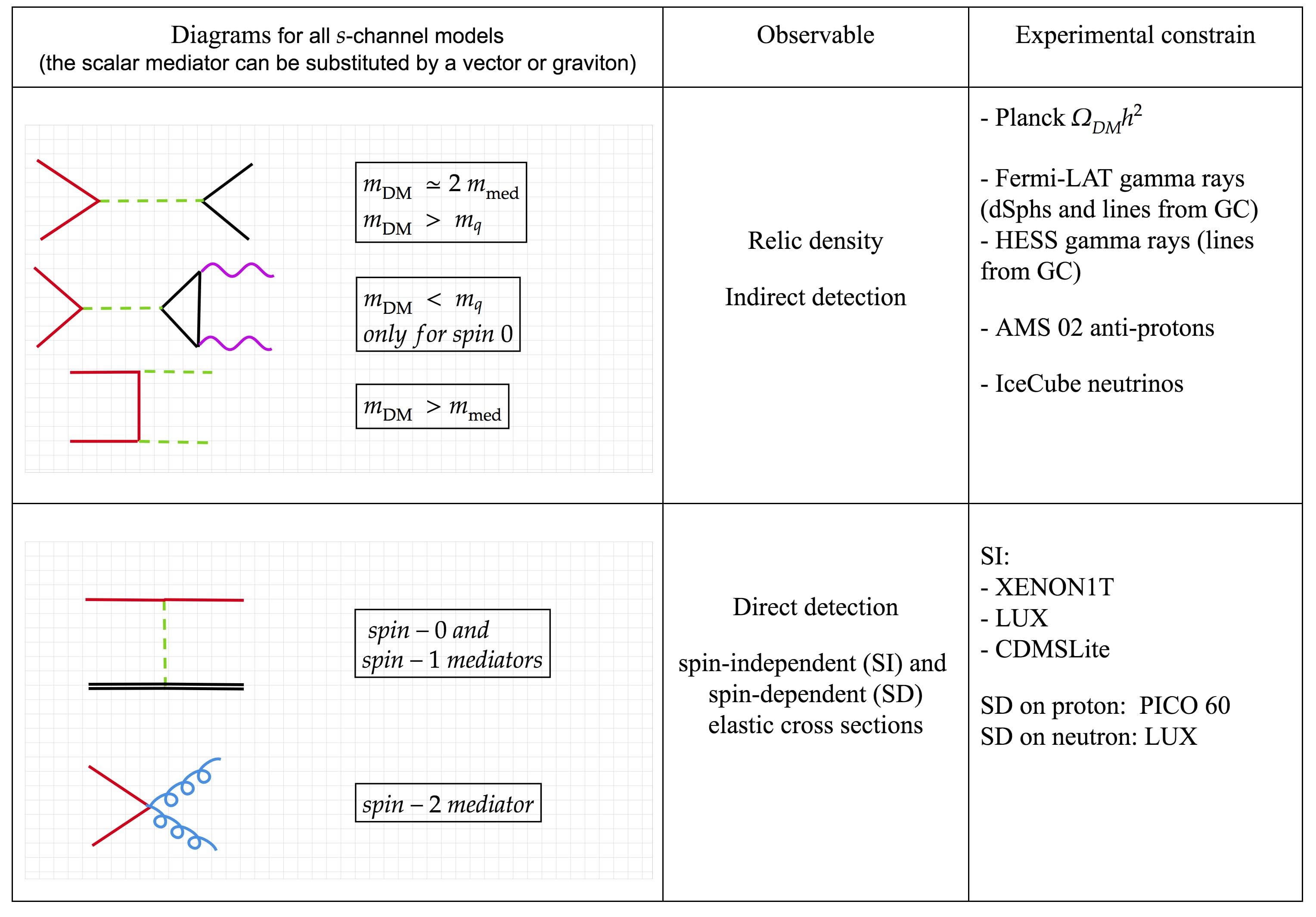}
\end{center}
\caption{Summary of the leading diagrams contributing to dark matter astrophysical and cosmological searches in the $s$-channel DMsimps. Colour code: red lines denote the dark matter particles, black lines are for SM fermions, purple lines for SM vector bosons, blue lines for gluons and green lines (dashed for the scalar mediator) for the mediator. The spin-1 and spin-2 cases are obtained simply by replacing $Y_0$ with the spin-1 and spin-2 mediators, except for the case of direct detection, where the first diagram contributes to SI and SD for spin-0 and spin-1, while the second diagram is for the spin-2 case in the minimalistic model. GC stands for Galactic Centre.}\label{fig:diagschans0s1}
\end{figure}

The material presented in this section is based on these selected reference papers~\citep{Arina:2016cqj,Haisch:2015ioa,Banerjee:2017wxi}, as they nicely exemplify the main features of scalar and pseudo-scalar mediators in the $s$-channel by performing comprehensive studies of the model, including astrophysical and cosmological dark matter searches.
\begin{figure}[t]
\begin{center}
\includegraphics[width=0.4\textwidth,trim=0mm 0mm 0mm 0mm, clip]{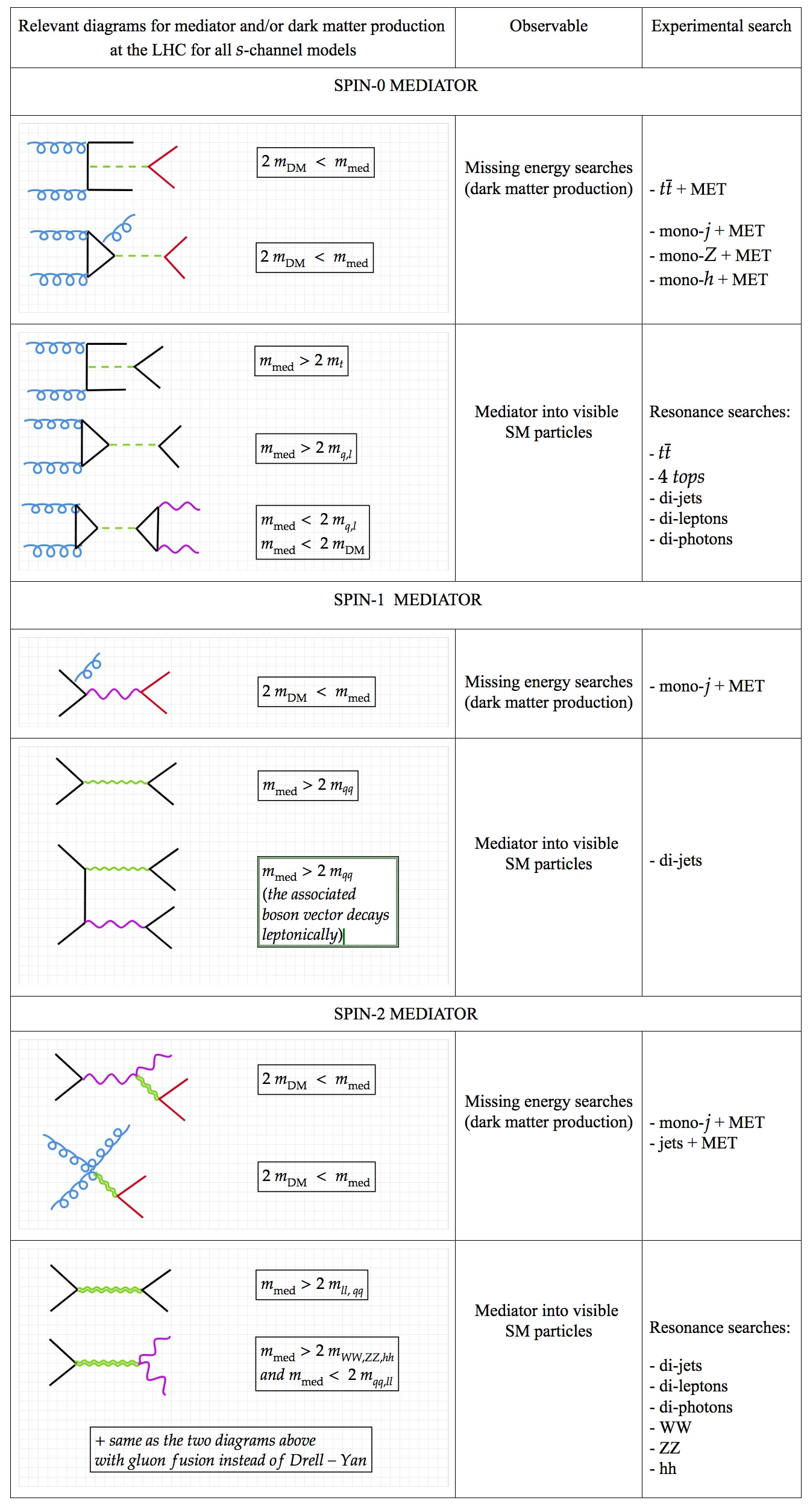}
\end{center}
\caption{Schematic of leading order diagrams contributing to dark matter and mediator searches at the LHC in the $s$-channel DMsimps. MET stands for missing transverse energy. The colour code is as in Fig.~\ref{fig:diagschans0s1}.}\label{fig:lhcschan}
\end{figure}

We focus on the case of Dirac dark matter ($X$), with spin-0 mediator ($Y_0$) coupling to the matter fields of the SM (the dependence on the dark matter spin is briefly summarised in Tab.~\ref{tab:Dmnat}).
The interaction Lagrangians is defined as:
\begin{equation}\label{eq:scalar_mediator}
 {\mathcal L}_{X}^{Y_0} = \bar{X} (\gdm^{S}+i \gdm^{P}\gamma_5)X\, Y_0 \,,
\end{equation}
and
\begin{equation}\label{eq:scalar_mediator2}
  {\mathcal L}_{\rm SM}^{Y_0}  =
   \sum_{i,j} \Big[\bar{d_i} \frac{y_{ij}^d}{\sqrt{2}}
       (g^{S}_{d_{ij}}+i g^{P}_{d_{ij}}\gamma_5)d_j  + \bar{u_i} \frac{y_{ij}^u}{\sqrt{2}}
       (g^{S}_{u_{ij}}+i g^{P}_{u_{ij}}\gamma_5)u_j\Big] Y_0 \,,
\end{equation}
where $d$ and $u$ denote down- and up-type quarks, respectively, 
($i,j$=1,2,3) are flavour indices, $\gdm^{S/P}$ are the scalar/pseudo-scalar WIMP-$Y_0$ couplings.  Following the prescriptions of MFV, the couplings of the mediator to the SM particles are proportional to the
particle masses and normalised to the SM Yukawa couplings, $y_{ii}^f = \sqrt{2}m_f/v$ and $v$ being the Higgs vacuum expectation value, and all flavour off-diagonal couplings are set to zero. 

The pure scalar and pure pseudo-scalar mediator scenarios, which we will review in the rest of the section, are given by 
setting the parameters in the Lagrangians~\eqref{eq:scalar_mediator} and \eqref{eq:scalar_mediator2} to:
\begin{align}
 & \gdm^S \equiv\gdm \quad {\rm and}\quad \gdm^P = 0
 \label{paramX_s}\,, \\
 & g^{S}_{u_{ii}}=g^{S}_{d_{ii}} \equiv \gsm \quad {\rm and}\quad
  g^{P}_{u_{ii}} =g^{P}_{d_{ii}} = 0\,,
 \label{paramSM_s}
\end{align}
and
\begin{align}
 & \gdm^S = 0 \quad {\rm and}\quad \gdm^P \equiv \gdm
 \label{paramX_p}\,,\\
 & g^{S}_{u_{ii}} = g^{S}_{d_{ii}} = 0 \quad {\rm and}\quad
  g^{P}_{u_{ii}} = g^{P}_{d_{ii}} \equiv \gsm \,,
 \label{paramSM_p}
\end{align}
respectively. With the simplification of a single universal coupling for the SM-$Y_0$ interactions, the model has only four independent parameters, \ie two couplings and two masses:  
\begin{equation}\label{param}
 \{g_{\rm SM},\,\gdm,\, \mdm,\, \mmed\} \,.
\end{equation}
The MFV assumption implies that we can even further neglect the contributions of all quarks but the top-quark in the model, as it has the largest Yukawa coupling. This is certainly an optimal approximation for LHC studies, while dark matter searches are sensitive to all quark flavours. The assumption  however that $\gsm \equiv g^{S/A}_{u_{33}}$ provides a good description of the phenomenology of the model, as the inclusion of all other quark flavours has the effect of globally decreasing the value of $\gsm$ needed to achieve the same cross section.

The Lagrangians of Eqs.~\eqref{eq:scalar_mediator} and~\eqref{eq:scalar_mediator2} induce dimension-five couplings of
the mediator to gluons and photons via top-quark loop diagrams.  These loop-induced
operators are relevant for both astrophysical and collider searches for dark matter. For a scalar $Y_0$, the couplings of the mediator to gluons and photons
are given, at the leading order, by the effective operators:
\begin{equation}
	{\mathcal L}^{\y}_{g} =-\frac{1}{4} \frac{\gglu(Q^2)}{v}\, G^a_{\mu\nu} G^{a,\mu\nu} \y
        \qquad\text{and}\qquad
	{\mathcal L}^{\y}_{\gamma} =-\frac{1}{4} \frac{\ggam(Q^2)}{v}\, F_{\mu\nu} F^{\mu\nu} \y\,,
\end{equation}
with the effective couplings being
\begin{equation}
	\gglu(Q^2) = \gsm \frac{\alpha_{s}}{3\pi}\,\frac{3}{2}  F_S \Big( \frac{4 m^2_t}{Q^2}\Big)
        \qquad\text{and}\qquad
	\ggam(Q^2) = \gsm \frac{8\alpha_{e}}{9\pi}\,\frac{3}{2} F_S \Big( \frac{4 m^2_t}{Q^2}\Big)\,, \label{eq:couplings}
\end{equation}
where $Q^2$ denotes the virtuality of the $s$-channel resonance, while $F_S$ is  the one-loop form factor
\begin{equation}\label{eq:fsg}
F_S(x) = x\Big[1 + (1-x) \, {\rm \arctan}^2 \Big( \frac{1}{\sqrt{x-1}} \Big)\Big]\,.
\end{equation}
Similar expressions can be retrieved for the pseudo-scalar case, see \eg~\citep{Haisch:2015ioa,Arina:2017sng}. Because of the hierarchy between the strong and the electromagnetic couplings ($\alpha_s^2/\alpha_e^2 \sim 100$), the $\y$ partial width into a pair of gluons is always larger than the one into a pair of photons. The expressions for tree level and loop-induced partial widths are provided in Ref.~\citep{Arina:2016cqj}.

Let us first discuss the case of pure scalar $\y$ and summarise briefly all the relevant LHC and dark matter searches to constrain its parameter space:
\begin{itemize}
\item {\bfseries LHC $\MET$ searches.} As this DMsimp features Yukawa-type couplings, the most relevant tree-level process at the LHC is dark matter pair production associated with a top-quark pair~\citep{CMS:2014pvf}. Similarly to Higgs production, at one loop, gluon fusion gives rise to $\MET$ + jet signatures~\citep{Khachatryan:2014rra}, mono-$Z$~\citep{Khachatryan:2015bbl} and mono-$h$~\citep{Aad:2015dva}, which are phenomenologically relevant.
\item {\bfseries LHC mediator searches.} The mediator is produced in association with top-quark pairs~\citep{Aad:2015fna}, or via the loop-induced gluon fusion process. These searches are relevant for mediators produced on-shell, or close to on-shell, which decay back into top pairs if kinematically allowed, or a pair of jets~\citep{CMS:2015neg} or photons~\citep{Khachatryan:2015qba}. For the heavy mediator case, the four-top final state~\citep{Khachatryan:2014sca} can be also relevant.
\item {\bfseries Relic density.} The dark matter achieves the correct relic density in three separated regions. If $\mdm > \mmed$ the relic density is set by the $t$-channel annihilation into a pair of mediators. Above the top threshold, resonant annihilation into top-quark pairs is efficient enough to lead to the correct value for $\relic$. For $\mdm < m_t$ the resonant annihilation into a pairs of gluon leads to the correct relic density for a very fine tuned part of the parameter space. This is due to the very small decay width into gluons.
\item {\bfseries Indirect detection.} All annihilation processes are $p$-wave suppressed, hence all fluxes of gamma rays, cosmic rays and neutrinos produced by this model are well below the present and future reach of indirect detection probes.
\item {\bfseries Direct detection.} The interaction Lagrangians in Eqs.~\eqref{eq:scalar_mediator} and~\eqref{eq:scalar_mediator2}, after some manipulations to express it in terms of nucleus instead of nucleons, reduces to the operator $\bar{X} X \bar{N} N$. This is equivalent to the $\mathcal{O}_1^{\rm NR}$ operator (see Tab.~\ref{tab:DDeft}), which corresponds to the usual spin-independent interaction. The scalar DMsimp is hence highly constrained by the XENON1T and LUX experimental upper bounds.
\end{itemize}
All the leading order relevant diagrams for $\y$ and dark matter production at the LHC and dark matter annihilation/scattering in astroparticle experiments are summarised in Figs.~\ref{fig:diagschans0s1} and~\ref{fig:lhcschan}.
%
\begin{figure}[t]
\begin{center}
\includegraphics[width=1.\textwidth,trim=0mm 0mm 0mm 0mm, clip]{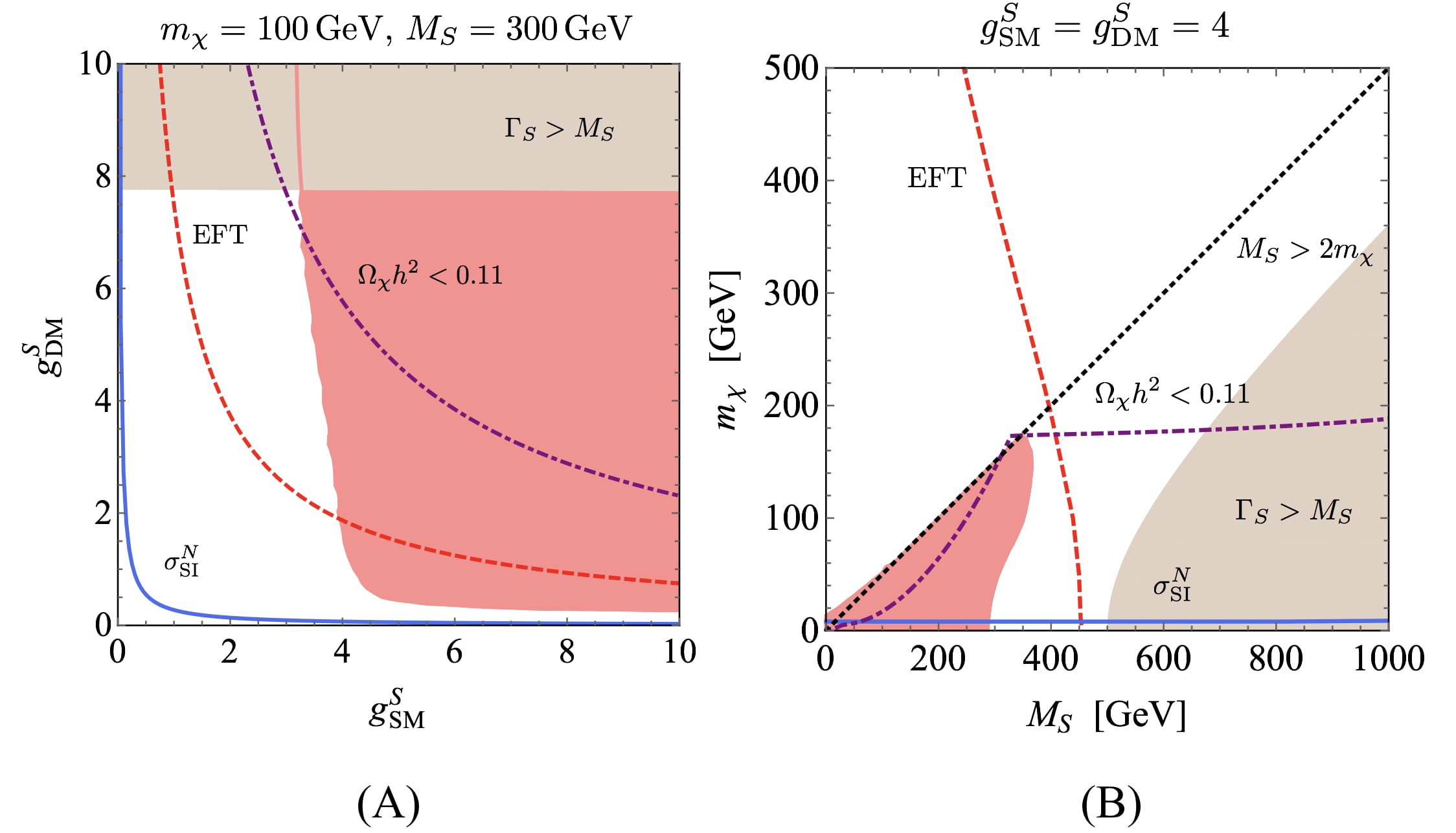}
\end{center}
\caption{{\bfseries DMsimp: $s$-channel spin-0 scalar mediator and Dirac dark matter.} Panel (A): Present mono-jet exclusion region at 95\% CL (red contour and region within) for scalar mediators in a 2D scan of the parameter space in the $\{\gdm^S, \gsm^S\}$-plane. The fixed values of the two parameters over which the scan is not performed are indicated in each panel.  For comparison, we show the region $\Gamma_S > M_S$ (brown, with $\Gamma_S$ being the mediator width), the LUX 90\% CL exclusion limits on $\sigma^{\rm SI}_n$ (solid blue curve, excludes above and on the right of the curve), the parameter space for under-abundant dark matter ($\relic < 0.11$, dot-dashed purple line), the EFT limit (red dashed line) and the region for which $M_S > 2 m_\chi$ (black dotted line). Panel (B): Same as (A) in the $\{m_\chi,M_S \}$-plane. Figures taken from Ref.~\citep{Haisch:2015ioa}. The reader can identify $\gdm^S = \gdm$, $\gsm^S = \gsm$, $m_\chi = \mdm$ and $M_S = \mmed$ with respect to the convention used in the review.} \label{fig:haischre}
\end{figure}
\begin{figure}[t]
\begin{center}
\includegraphics[width=1.\textwidth,trim=0mm 0mm 0mm 0mm, clip]{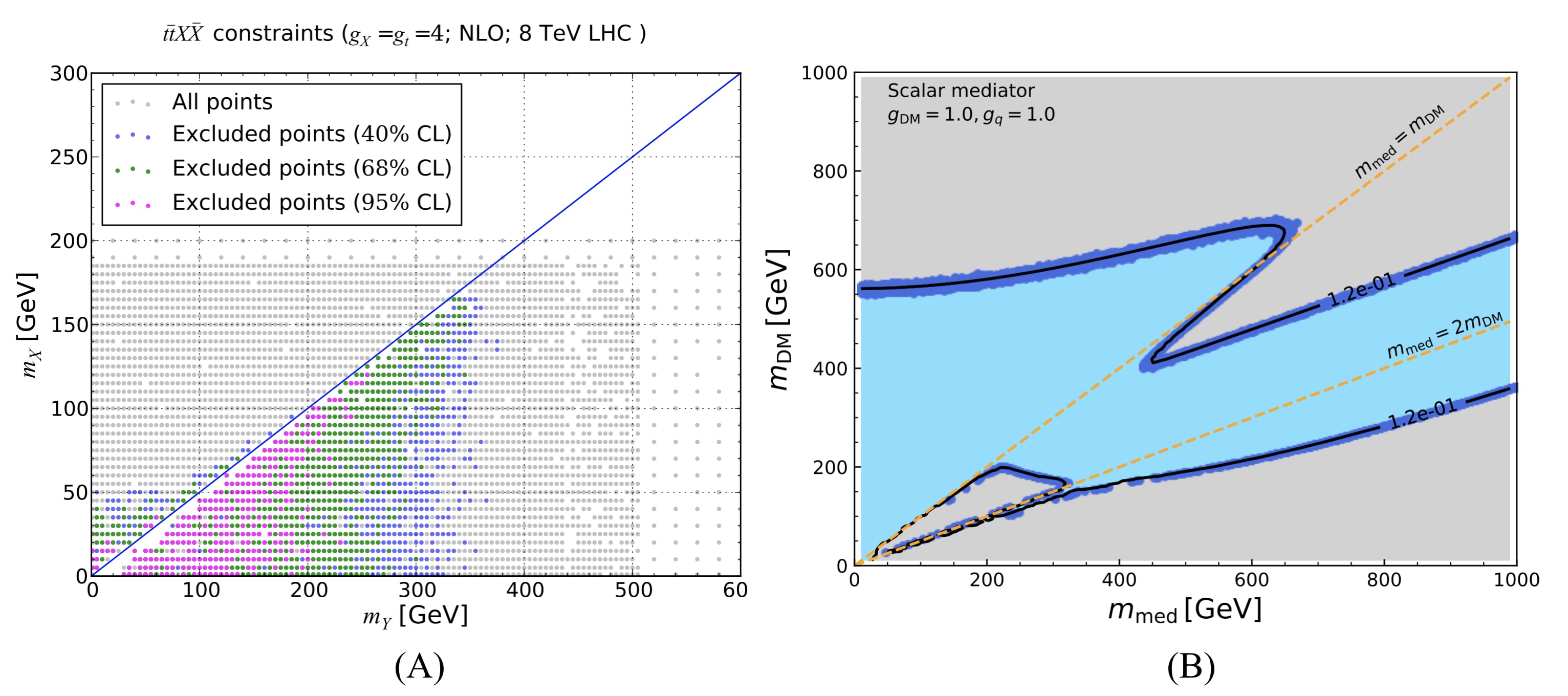}
\end{center}
\caption{{\bfseries DMsimp: $s$-channel spin-0 scalar mediator and Dirac dark matter.} Panel (A):  Constraints from the CMS 8TeV $t\bar{t}+\MET$ analysis in the $\{m_X, m_Y\}$-plane. The top and dark matter couplings to the mediator are set to 4, as labelled. The next to leading order (NLO) exclusions are shown. Figure taken from~\citep{Arina:2016cqj}. The reader can identify $m_X =\mdm$ and $m_Y = \mmed$ with respect to the convention used in the review. Panel (B):  Dark matter relic density in the $\{ \mdm, \mmed \}$-plane. The grey region denotes over-abundant dark matter, while the light blue region is for under-abundant dark matter.  The black solid line/dark blue points denote the parameter space for which the dark matter has the correct relic density. The orange dashed lines stand for $\mmed = \mdm$ and $\mmed = 2 \mdm$, as labelled. The couplings are fixed at the values labelled in the plot. Figure taken from~\citep{Ambrogi:2018jqj}. The reader can identify $g_q = \gsm$ with respect to the convention used in the review.} \label{fig:my1}
\end{figure}
\begin{figure}[t]
\begin{center}
\includegraphics[width=0.6\textwidth,trim=0mm 0mm 0mm 0mm, clip]{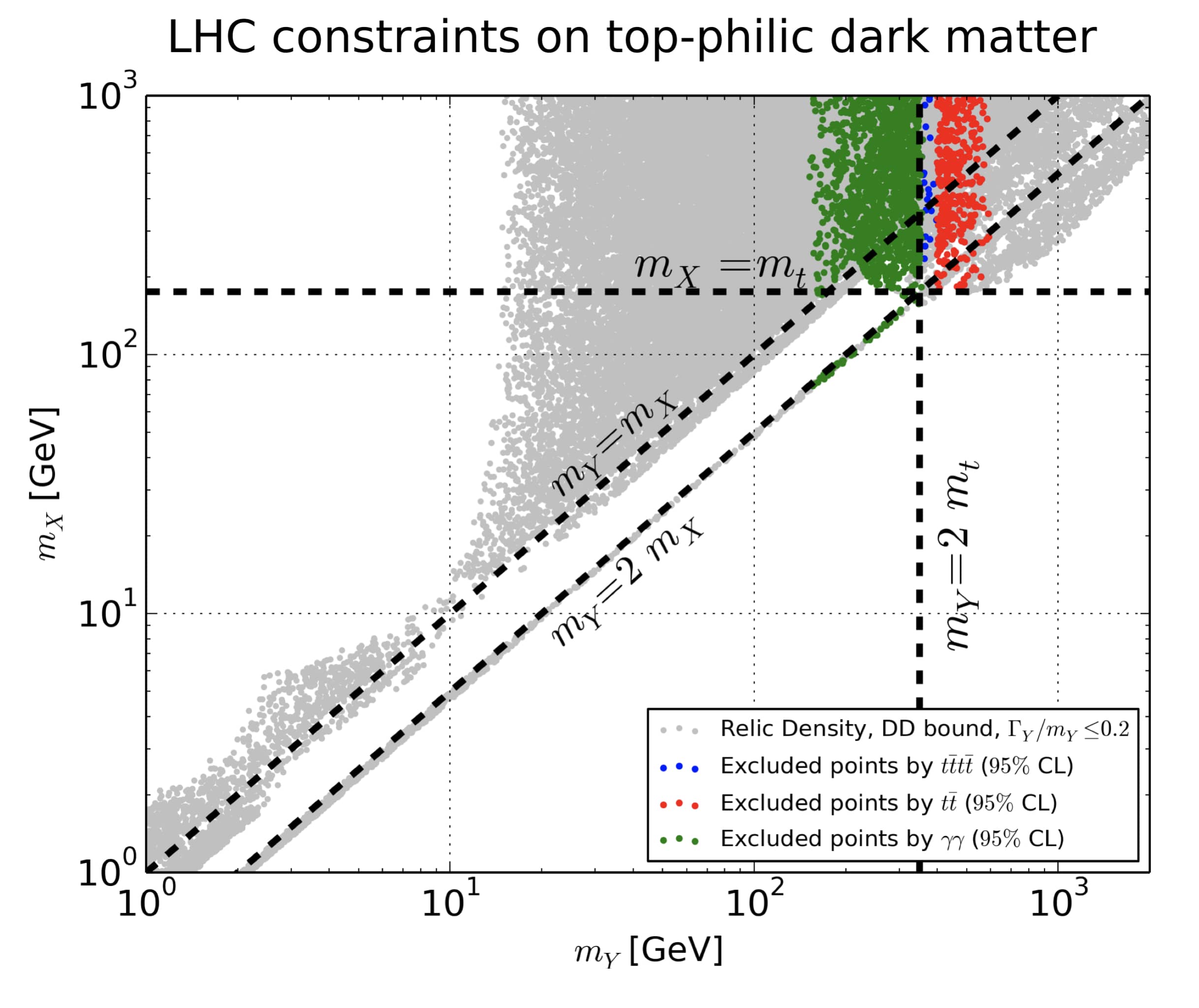}
\end{center}
\caption{{\bfseries DMsimp: $s$-channel spin-0 scalar mediator and Dirac dark matter.} Results of a 4D parameter sampling projected onto the  $\{m_X,m_Y\}$-plane, assuming a scenario of thermal dark matter. All grey points satisfy the relic density, narrow width assumption and direct detection constraints. The white region with $m_X < m_Y$ is excluded by relic density constraints (over-closure of the universe), while in the left upper corner the white region is excluded by the LUX and CDMSLite upper limits at 90\% CL. LHC constraints are imposed by the coloured points, as labelled. The green points are excluded by the di-photon searches, the red points are excluded by $t\bar{t}$ searches and the blue points by the four-top search. Figure taken from~\citep{Arina:2016cqj}. The reader can identify $m_X =\mdm$ and $m_Y = \mmed$ with respect to the convention used in the review.} \label{fig:my2}
\end{figure}

The result of the comprehensive studies are presented in Figs.~\ref{fig:haischre},~\ref{fig:my1} and~\ref{fig:my2}, from~\citep{Haisch:2015ioa,Arina:2016cqj,Ambrogi:2018jqj}, assuming a narrow width approximation. Figure~\ref{fig:haischre} illustrates the mono-jet + $\MET$ constraints on fixed slices of the model parameter space (red regions). It is clear that mono-jets + $\MET$ searches constrain the model parameter  space for large values of the $\y$-SM coupling, $\gsm \geq 3.5$. The same couplings contribute to the direct detection signal, $\sigma^{\rm SI}_n \propto \gsm^2 \gdm^2 / \mmed^4$, and lead to large elastic scattering cross sections, already excluded by LUX (blue solid line). Also shown is the EFT limit, which sets in for heavy mediators. Notice that mono-jets (and mono-X) + $\MET$ searches are sensitive to the region $\mmed > 2 \mdm$, where typically the dark matter over-closes the universe, if considered as a pure thermal relic. Figure~\ref{fig:my1}, panel (A), illustrates the reach of the $t\bar{t} + \MET$ search at 8 TeV, where NLO simulations, that reduce the theoretical errors, are used. Similarly to the case of mono-jets + $\MET$, the mediator should be heavier than twice the dark matter mass, in order to be able to decay into invisible states; and the constraints are sensitive to large $\gsm$ couplings. In panel (B) we show the behaviour of the relic density calculation for a 2D scan over the mass parameters and couplings fixed at 1 (this is one of the benchmark point recommended by the LHC DM working group~\citep{Boveia:2016mrp}). The black line represents the values of masses that achieve the correct $\relic$, the blue region denotes under-abundant dark matter (mostly leaving in the region $\mdm > \mmed$ and dominated by the $t$-channel annihilation into mediator pairs), while the grey region stands for over-abundant dark matter (mostly covering the region $\mmed > 2 \mdm$, where $\MET$ searches are relevant).
Figure~\ref{fig:my2} illustrates a comprehensive parameter space sampling of the model, with the assumption that the dark matter is a thermal relic and constitutes 100\% of the matter content of the universe. Couplings are freely varied in between $10^{-4}$ and $\pi$. The relic density measurement rules out completely the region sensitive to $\MET$ searches, while direct detection disfavours at 90\% CL regions with a light mediator for a wide range of $\mdm$. Resonance searches are relevant and constrain the region $\mdm > m_t$. Di-photons are sensitive to the parameter space $\mmed <  2m_t$, while the $t\bar{t}$ and 4 top searches are sensitive to $\mmed > 2 m_t$. A summary of the search sensitivities is provided in Tab.~\ref{tab:Dmnat}.

\begin{figure}[h]
\begin{center}
\includegraphics[width=1.\textwidth,trim=0mm 0mm 0mm 0mm, clip]{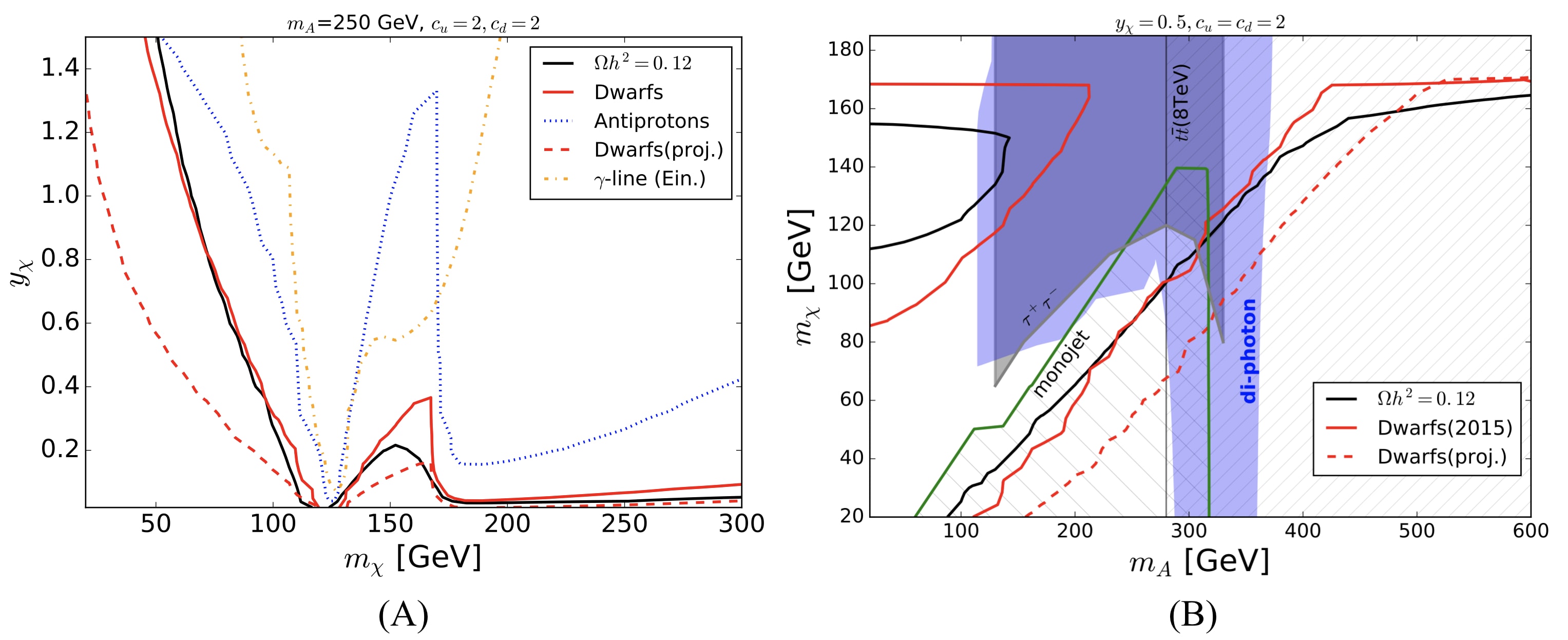}
\end{center}
\caption{{\bfseries DMsimp: $s$-channel spin-0 pseudo-scalar mediator and Dirac dark matter.} Panel (A): Dark matter constraints on the model parameter space in the $\{y_\chi, m_\chi \}$-plane. The other parameters are fixed as labelled. Below the black line the universe is over-closed, while the region above the red solid line is excluded by the Fermi-LAT dSph gamma-ray searches. The region above the dotted blue line is disfavoured by AMS 02 anti-proton measurements, whereas the region above the yellow dot-dashed line is excluded at 95\% CL by gamma-ray line searches from the Galactic Center. The red dashed curve denotes the expected sensitivity of the Fermi-LAT searches in dSPhs after 15 years of data. Panel (B): Dark matter and collider searches presented in the $\{m_\chi, m_A \}$-plane. The other parameters are fixed as labelled. If considered as thermal relic the dark matter allowed region is contained in between the red and black solid lines. The shaded regions are excluded by LHC searches at 95\% CL: mono-jet (hatched green), $A \to \tau^+ \tau^-$ (grey), di-photons (blue) and $t\bar{t}$ (hatched grey). Figures taken from~\citep{Banerjee:2017wxi}. The reader can identify $m_\chi =\mdm$ and $m_A = \mmed$, $c_u=c_d=\gsm$ and $y_\chi=\gdm$ with respect to the convention used in the review.} \label{fig:pseudo}
\end{figure}

Moving on to the pure pseudo-scalar case~\citep{Banerjee:2017wxi}, the relevant LHC and dark matter searches are:
\begin{itemize}
\item {\bfseries LHC $\MET$ searches.} These are the same as for the scalar case.
\item {\bfseries LHC mediator searches.} These are the same as for the scalar case. By switching on the couplings to leptons, an additional relevant search is the production via gluon fusion or in association with a pair of bottom-quarks, of the mediator decaying into a pair of $\tau$ leptons ($A \to \tau^+ \tau^-$)~\citep{CMS:2016rjp}. This holds for a scalar $Y_0$ as well.
\item {\bfseries Indirect detection.} The annihilation channels with $\y$ exchanged in the $s$-channel are $s$-wave dominated (\ie $X \bar{X} \to gg, t\bar{t}$), hence the pseudo-scalar mediator model can be constrained by gamma-ray and cosmic-ray searches.
\item {\bfseries Direct detection.} Direct detection is not sensitive to pseudo-scalar mediators. This can be understood by looking at Tab.~\ref{tab:DDeft}: the high-energy Lagrangians of the pure pseudo-scalar case are mapped into $\mathcal{O}^{NR}_6$. This non-relativistic operator is suppressed by the momentum transfer to the fourth power, hence the current direct detection experiments are insensitive to it, unless the mediator is of the order of the MeV~\cite{Arina:2014yna}. 
\end{itemize}
The result of the analysis are illustrated in Fig.~\ref{fig:pseudo} from~\citep{Banerjee:2017wxi}. Panel (A) shows all astrophysical and cosmological constraints for the dark matter: Fermi-LAT exclusion limits from dSphs are more stringent than both anti-proton bounds (as well as more robust in terms of astrophysical uncertainties) and gamma-ray line searches. Panel (B) shows the most stringent dark matter constraints combined with the LHC searches. A thermal relic scenario lives in the narrow band in between the black and the red solid lines. It is a narrow region because it is dominated by resonant s-channel annihilation, which is fine tuned however occurs in all dark matter models featuring an $s$-channel mediator. $\MET$ searches probe a region which is already challenged by the  Fermi-LAT dSph constraints. On the other hand, di-photons, $t\bar{t}$ and $\tau$ leptons can probe the mediator mass as low as 100 GeV and challenge the left-hand side region where dark matter is a viable thermal relic. The projection for the exclusion bounds coming from the Fermi-LAT satellite after 15 years of operation (red dashed line) shows that these data can basically probe the whole parameter space of the model (everything on the left hand side of the curve is excluded). Notice that additional dark states and mediators can affect the relic density and indirect detection regions. However the changes are supposed to go both in the same directions, hence the region allowed by Planck and Fermi-LAT will remain narrow.  LHC bounds for $\mmed < 2 m_t$ can change sensibly if additional scalars are introduced, as new decay channels will become available; conversely the constraints for $\mmed > 2 m_t$ are robust and will be qualitatively unaltered.

Other studies of the spin-0 case are for instance~\citep{Harris:2014hga,Buckley:2014fba,Pree:2016hwc,Dolan:2016qvg}, while details on loop-induced process for mono-jet + MET can be found in Refs.~\cite{Haisch:2012kf,Buckley:2014fba,Harris:2014hga,Haisch:2015ioa,Backovic:2015soa}. Leptonic couplings have been introduced in \eg~\citep{Albert:2017onk}. Similarly, $\y$ couplings to the SM gauge bosons are discussed in~\cite{Neubert:2015fka}.

\subsubsection{Spin-1 mediator}\label{sec:spin1}

The material discussed in this section is based on these selected Refs.~\citep{Chala:2015ama,Heisig:2015ira,Carpenter:2016thc,Albert:2017onk}, that exhaustively exemplify the main features of vector and axial-vector mediators in the $s$-channel and perform comprehensive studies of the model, including astrophysical and cosmological dark matter searches.

The interaction Lagrangian of a spin-1 mediator ($Y_1$) with a Dirac fermion dark matter particle ($X$) is given by:
\begin{equation}\label{eq:vector_mediator}
 {\mathcal L}_{X}^{Y_1} = \bar{X} \gamma_{\mu} ({\gdm^{V}}+{\gdm^{A}}\gamma_5)X\,Y_1^{\mu} \,,
\end{equation}
and with quarks by:
\begin{equation}\label{eq:vector_mediator2}
  {\mathcal L}_{\rm SM}^{Y_1} = \sum_{i,j} \Big[\bar{d_i} \gamma_{\mu}
    (g^{V}_{d_{ij}}+g^{A}_{d_{ij}}\gamma_5)d_j  +\bar{u_i}  \gamma_{\mu}
    (g^{V}_{u_{ij}}+g^{A}_{u_{ij}}\gamma_5)u_j\Big] Y_1^{\mu} \,,
\end{equation}
where ${\gdm^{V/A}}$ and $g^{V/A}_{u/d_{ij}}$ are the vector/axial-vector couplings of the dark matter and quarks with $Y_1$. For a Majorana dark matter candidate the vector coupling is not allowed.

The pure vector and pure axial-vector mediator scenarios are obtained by
setting the parameters in the Lagrangians~\eqref{eq:vector_mediator} and 
\eqref{eq:vector_mediator2} to
\begin{eqnarray}
 &g^V_{X} \equiv \gdm \quad {\rm and}\quad g^A_{X} = 0
 \label{paramX_v}\,, \\
 &g^{V}_{u_{ii}} =  g^{V}_{d_{ii}} \equiv \gsm \quad {\rm and}\quad
  g^{A}_{u_{ii}} =  g^{A}_{d_{ii}} = 0
 \label{paramSM_v}
\end{eqnarray}
and 
\begin{eqnarray}
 &g^V_{X} = 0 \quad {\rm and}\quad g^A_{X_D} \equiv \gdm
 \label{paramX_a}\,,\\
 &g^{V}_{u_{ii}} = g^{V}_{d_{ii}} = 0 \quad {\rm and}\quad
  g^{A}_{u_{ii}} = g^{A}_{d_{ii}} \equiv \gsm\,,
 \label{paramSM_a}
\end{eqnarray}
respectively, where we assume quark couplings to the mediator to be flavour universal and set all flavour off-diagonal couplings to zero. Similarly to the case of spin-0 mediator, this model has only four free parameters, defined as in Eq.~\eqref{param}. The universality assumption of the couplings is also justified by gauge invariance, which sets very tight constraints on the relation among couplings, see \eg~\citep{Bell:2015sza}. Even though the Lagrangians presented above do not preserve gauge invariance, the assumption of having different couplings to up- and down-type quarks, as \eg in~\citep{Chala:2015ama}, can lead to artificial enhanced cross sections which are not representative of gauge invariant theories. 

In this model the couplings to leptons are not considered, hence it can be seen as a lepto-phobic $Z'$ model, see \eg~\citep{Duerr:2014wra}. Leptonic couplings are indeed very tightly constrained by di-lepton resonant searches~\citep{Dudas:2009uq,Arcadi:2013qia,Lebedev:2014bba} and can be switched off to allow to have large quark couplings.

Let us first discuss the complementarity of searches for the case of a pure vectorial $Z'$ model, hence the dark matter candidate can only be a Dirac fermion~\citep{Chala:2015ama,Pree:2016hwc,Carpenter:2016thc}. 
\begin{itemize}
\item {\bfseries LHC $\MET$ searches.}  ATLAS and CMS searches for jets in association with $\MET$ (due to initial state radiation of a gluon) place strong constraints on this model~\citep{Khachatryan:2014rra,Aad:2015zva}. 
\item {\bfseries LHC mediator searches.} The di-jet final state is a very important  complementary channel, as it has been pointed out in~\citep{Chala:2015ama}. Di-jets can be produced via $Y_1$ Drell-Yan process or via associated production. Stringent bounds for di-jet invariant mass above 1 TeV are provided by ATLAS~\citep{Aad:2014aqa,TheATLAScollaboration:2013gia} and CMS~\citep{Khachatryan:2015sja}, while complementary and equally tight bounds for smaller masses are provided by the UA2~\citep{Alitti:1993pn} experiment and the Tevatron CDF experiment~\citep{Aaltonen:2008dn}.
\item {\bfseries Relic density.} The dark matter achieves the correct relic density in a small narrow band for fixed couplings. If $\mdm > \mmed$ the relic density is set by the $t$-channel annihilation into a pair of mediators, which is an $s$-wave process proportional to $\gdm^4$. For $\gdm \sim 1$ this cross section is small and the dark matter is under-abundant. For the benchmark points chosen by the LHC Dark matter working group~\citep{Albert:2017onk}, the correct relic density is achieved by the exchange in the $s$-channel of a $Y_1$, leading to resonant annihilation into quark pairs, which is also $s$-wave. Of course, the introduction of leptonic couplings can change this classification. 
\item {\bfseries Indirect detection.} $\sigmav_0$ receives contributions from the same channels that fix the relic density. For the details on the annihilation cross section we refer to~\citep{Albert:2017onk}. However in the literature, at the best of our knowledge, there are no results on constraints from Fermi-LAT dSph gamma-ray searches that include the $t$-channel term.
\item {\bfseries Direct detection.} The interaction Lagrangians in Eqs.~\eqref{eq:vector_mediator} and~\eqref{eq:vector_mediator2} are equivalent to $\mathcal{O}_1^{\rm NR}$, see Tab~\ref{tab:DDeft}. This non-relativistic operator describes the usual spin-independent elastic scattering off nuclei. The vector model is hence highly constrained by the XENON1T and LUX experimental upper bounds.
\end{itemize}
The leading order relevant diagrams for $Y_1$ and dark matter production at the LHC and dark matter annihilation/scattering in astroparticle experiments are summarised in Figs.~\ref{fig:diagschans0s1} and~\ref{fig:lhcschan} (the same holds for the pure axial-vector mediator). 
%
\begin{figure}[t]
\begin{center}
\includegraphics[width=1.\textwidth,trim=0mm 0mm 0mm 0mm, clip]{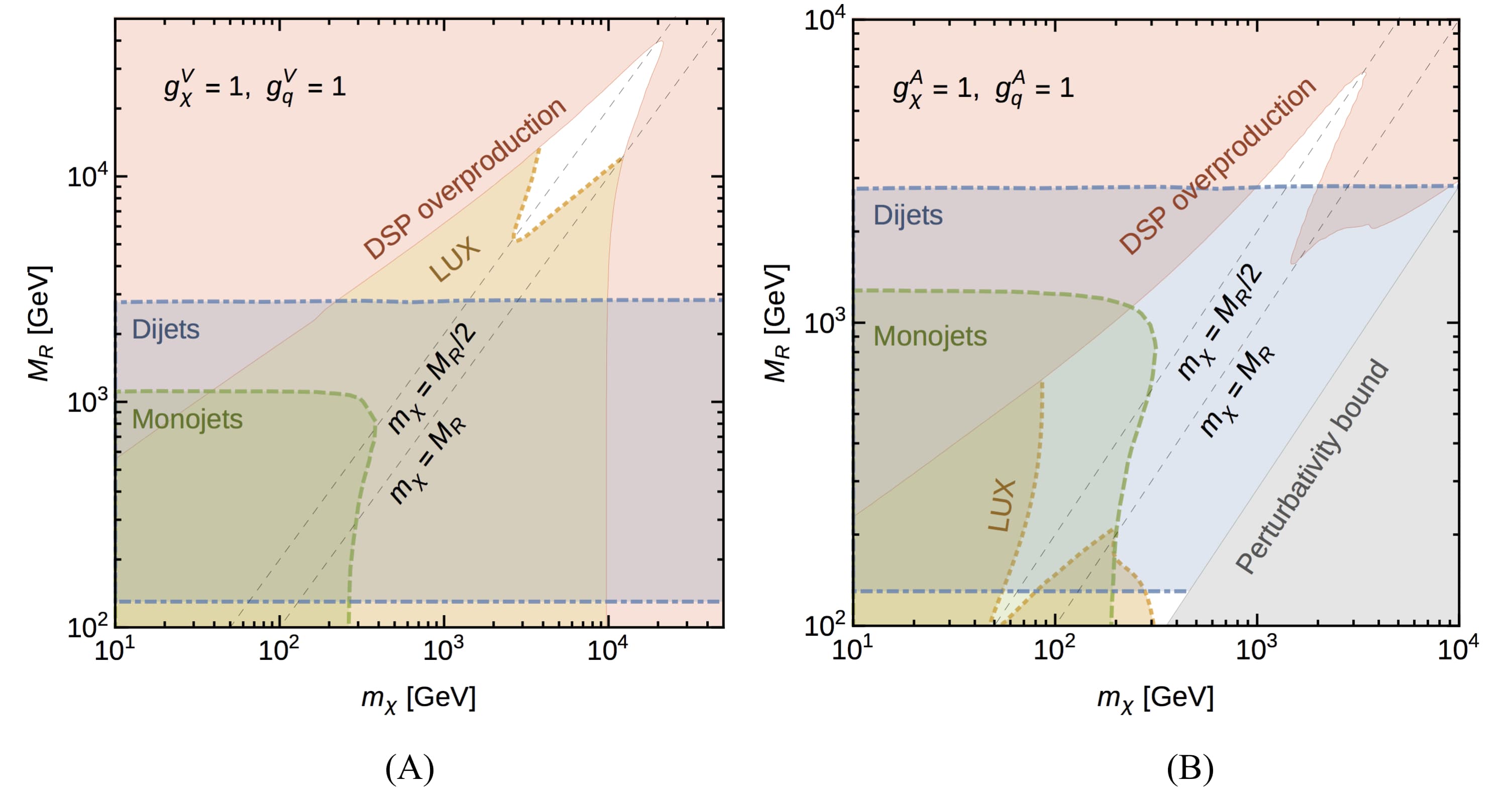}
\end{center}
\caption{{\bfseries DMsimp: $s$-channel spin-1 vector (panel (A)) and axial-vector (panel (B)) mediator and Dirac dark matter.} Panel (A): Combined constraints at 95\% CL from the LUX experiment (orange dotted line and orange shaded region), from mono-jet searches (green dashed line and green shaded region) and di-jets (blue dot-dashed line and region in between) in the $\{ M_R, m_\chi\}$-plane for fixed couplings, as labelled. We also show the region that over-closes the universe (red) and the region excluded by perturbativity (grey). Panel (B):  Same as panel (A). Figures taken from~\citep{Chala:2015ama}. The reader can identify $m_\chi =\mdm$ and $M_R = \mmed$, $g_\chi^{V/A}=\gdm$ and $g_q^{V/A}=\gsm$ with respect to the convention used in the review.} \label{fig:avdijets}
\end{figure}
\begin{figure}[t]
\begin{center}
\includegraphics[width=1.\textwidth,trim=0mm 0mm 0mm 0mm, clip]{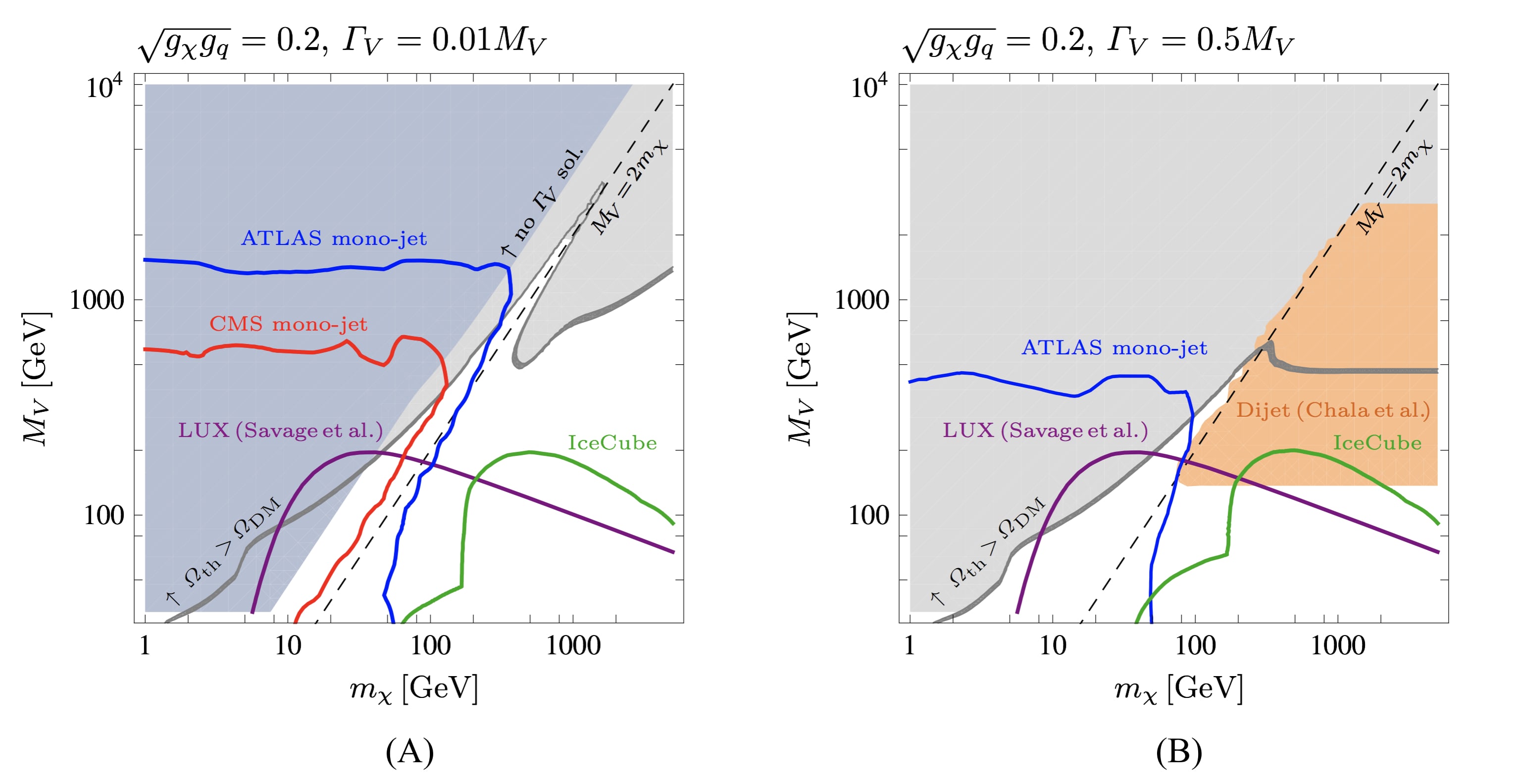}
\end{center}
\caption{{\bfseries DMsimp: $s$-channel spin-1 axial-vector mediator and Majorana dark matter.} Panel (A): Combined constraints in the $\{ M_V, m_\chi\}$-plane for fixed couplings and for a narrow mediator width $\Gamma_V = 0.01 M_V$, as labelled. We show the region disfavoured by mono-jet searches from ATLAS (blue solid line including the region below and on the left) and CMS (red solid line including the region below and on the left), both at 95\% CL, the LUX exclusion bound (purple solid line and region below) and by IceCube searches (green solid line and region below it), both at 90\% CL.The light grey region stands for over-abundant dark matter, while the grey band denotes the region where the dark matter relic density matches the observed one. The blue region does not allow for a consistent solution of $\Gamma_V$ in terms of $M_v$, $m_\chi$ and $\sqrt{g_\chi g_q}$ within this DMsimp. Panel (B):  Same as panel (A) for a large mediator width $\Gamma_V = 0.5 M_V$. The orange region denotes the constraint from di-jets searches. Figures taken from~\citep{Heisig:2015ira}. The reader can identify $m_\chi =\mdm$ and $M_V = \mmed$, $g_q=\gsm$ and $g_\chi=\gdm$ with respect to the convention used in the review.} \label{fig:majic}
\end{figure}

Panel (A) of Fig.~\ref{fig:avdijets}, from~\citep{Chala:2015ama}, shows the complementarity of collider, cosmological and direct detection searches, with fixed couplings $\gsm=\gdm=1$. Basically the whole parameter space of the model is strongly disfavoured by the current limits of direct detection experiments. Di-jets and mono-jets have a rather smaller impact on the model parameter space. Notice however that, contrary to the case of spin-0 mediator, collider searches are sensitive to smaller values of $\gsm$, even of the order of $\mathcal{O}(0.1)$. Mono-X searches are more sensitive to the region for which $\mmed > 2 \mdm$, in which the DMsimp features  over-abundant dark matter. This assumption can be circumvented by invoking for instance dark matter non thermal production or entropy injection. Conversely, di-jet constraints are rather independent of the dark matter mass and cover all dark matter regions.  Constraints from Fermi-LAT dSphs have been discussed in~\citep{Carpenter:2016thc}: the parameter space of the model is most restricted for $\mmed \sim 2 \mdm$, because of the enhancement in $\sigmav_0$ due to the resonance. If the vector mediator is much heavier than the dark matter, the total annihilation cross section drops and the parameter space becomes suddenly less constrained.  This can be understood by the fact that annihilation occurs far away from the resonance, hence $\sigmav_0$ decreases quickly.

Moving to the axial-vector case, the dark matter can be either Dirac or Majorana. The most relevant dark matter searches are~\citep{Chala:2015ama,Heisig:2015ira,Pree:2016hwc}:
\begin{itemize}
\item {\bfseries LHC $\MET$ searches and mediator searches.} These are exactly the same as in the pure vector case described above.
\item {\bfseries Relic density.} The $s$-channel process is helicity suppressed if $\gdm^V=0$, namely it is proportional to $m^2_q$, while the $t$-channel is $s$-wave, taken properly into account in the analysis in~\citep{Albert:2017onk}.
\item {\bfseries Indirect detection.} In the analyses performed so far there are no bounds from gamma-ray or cosmic-ray searches because the $t$-channel process has not been properly taken into account. However relevant constraints for the model parameter space arise from the IceCube upper limits on $\sigma^{\rm SD}_p$, where all annihilation processes contributing to $\sigmav_0$ have been properly taken into account.
\item {\bfseries Direct detection.} Spin-independent elastic scattering is superseded by the ordinary spin-dependent elastic scattering (corresponding to $\mathcal{O}_4^{\rm NR}$ in Tab.~\ref{tab:DDeft}). This operator is less constrained by direct detection experiments with respect to $\mathcal{O}_1^{\rm NR}$. The most constraining experiment is LUX for $\sigma^{\rm SD}_n$.
\end{itemize}

The right panel of Fig.~\ref{fig:avdijets}, from~\citep{Chala:2015ama}, describes the complementarity of collider, cosmological and direct detection searches, with fixed couplings $\gsm=\gdm=1$, for the axial-vector model. The impact of the LUX exclusion limit is rather reduced with respect to the pure vector case. Hence collider bounds have a nice degree of complementarity for this model, disfavouring the majority of the viable parameter space. The grey region is excluded by the perturbativity bound, obtained by imposing $\mmed > \gdm^4 \mdm \sqrt{4 \pi}$, which comes from the requirement that the annihilation cross section remains well-behaved at large dark matter masses. Figure~\ref{fig:majic}, from~\citep{Heisig:2015ira}, shows the impact of the IceCube bounds on the model parameter space for fixed product of the couplings and for a narrow $Y_1$ width (panel (A)) and for a large mediator width (panel (B)), as $\Gamma_{Y_1} \equiv \Gamma_V$ is taken as a free parameter. In the very narrow width approximation, di-jet constraints are irrelevant, while mono-jet + $\MET$ searches are much less affected by changes in the mediator width. The exclusion bound stemming from LUX does not depend on the mediator width, and remains unchanged in the two cases and constrain the DMsimp parameter space where dark matter is either a thermal relic or under-abundant. The IceCube exclusion limit has a subtle dependence on the mediator width, as the annihilation rate is sensitive to both the $s$-channel process, which depends on $\gsm \times \gdm$, and on the $t$-channel process, which depends only on $\gdm$, for $\mdm \geq \mmed$.\footnote{The exclusion bounds are not rescaled, as the authors assume that the dark matter makes up 100\% of the matter content of the universe in the white region. Thermal production is then supplemented by some other mechanism to achieve the observed value of $\relic$.} In panel (A), IceCube and LUX probe a complementary region, in which $\mdm > \mmed$, with respect to LHC searches. LUX constraints are relevant at intermediate dark matter masses, while IceCube lower limits overtake all other constraints at large dark matter masses. In case of a large mediator width, the IceCube bound overlaps with the di-jet constraints. From a refined analysis on di-jets in~\citep{Fairbairn:2016iuf}, it has been shown that for $\mmed < 3 \TeV$ and $\Gamma_{Y_1} > 0.25 \, \mmed$, the collider constraints disfavour the possibility that the WIMP-quark interactions are responsible for setting the dark matter relic density. A summary of the search sensitivities and their dependency on the dark matter nature is provided in Tab.~\ref{tab:Dmnat}.

The LHC Dark Matter working group has suggested to consider leptonic couplings as well~\citep{Albert:2017onk}. These should be however at least one order of magnitude smaller than the mediator-quark couplings, to not completely exclude the model. Interestingly couplings to neutrinos would also be present because of gauge invariance requirements; these couplings will supply an additional $\MET$ channel with the consequences of enhancing certain mono-X + $\MET$ signals.

Other studies of the spin-1 DMsimps are for instance~\citep{Harris:2014hga,Buchmueller:2014yoa,Jacques:2015zha,Pree:2016hwc,Fairbairn:2016iuf,Jacques:2016dqz,Brennan:2016xjh,Bell:2016fqf}. The latter papers in the list already consider a gauge invariant completion of the $Z'$ model, instead of the DMsimp Lagangrians in Eqs.~\eqref{eq:vector_mediator} and~\eqref{eq:vector_mediator2}. This issue will be discussed in Sec.~\ref{sec:caveats}.

\subsubsection{Spin-2 mediator}\label{sec:spin2}

The material presented in this section is based on these selected Refs.~\citep{Kraml:2017atm,Lee:2014caa,Zhang:2016xtc}, as they exemplify the main features of a spin-2 mediator in the $s$-channel as compared with LHC searches and indirect detection searches. The literature on spin-2 mediator is rather reduced with respect to the spin-0 and spin-1 cases. Relevant works are provided by these Refs.~\citep{Garcia-Cely:2016pse,Dillon:2016tqp,Dillon:2016fgw,Yang:2017iqh,Zhu:2017tvk,Rueter:2017nbk}.

Even though the exchange of a graviton in the $s$-channel is not considered in the recommendations of the LHC Dark Matter working group~\citep{Boveia:2016mrp}, it entails several features in common with the DMsimp philosophy. It is possible to build a dark matter simplified model out of a gravity-mediated dark matter model proposed in~\citep{Lee:2013bua}, even though it requires a dedicated validation work, as such model is, in general, not renormalisable. This type of models have as well driven a lot of attention at the time of the 750 GeV excess in the di-photon channel, see \eg~\citep{Han:2015cty,Arun:2015ubr,Martini:2016ahj} and the references therein.

The definition of the model follows the approach of DMsimps. We consider dark matter particles which interact with the SM particles via an $s$-channel spin-2 mediator. 
The interaction Lagrangian of a spin-2 mediator ($Y_2$) with the dark matter ($X$) is given by~\cite{Lee:2013bua}:
\begin{equation}\label{y-x}
 {\mathcal L}_{X}^{Y_2} = -\frac{1}{\Lambda} g^{T}_{X}\,T^X_{\mu\nu} Y_2^{\mu\nu}\,,
\end{equation}
where $\Lambda$ is the scale parameter of the theory, $g^T_X$ is the coupling between $Y_2$ and the dark matter, and $T_{\mu\nu}^X$ is the energy--momentum tensor of the dark matter field. The energy--momentum tensors of the dark matter are:
\begin{eqnarray}\label{em-x}
 T_{\mu\nu}^{X_R} & = -\frac{1}{2}g_{\mu\nu}(
  \partial_{\rho}X_R\partial^{\rho}X_R - m^2_{X}X^2_R) +\partial_{\mu}X_R\partial_{\nu}X_R\,,\\ 
 T_{\mu\nu}^{X_D}  & = -g_{\mu\nu}(
  \overline{X}_Di\gamma_{\rho}\partial^{\rho}X_D - m_{X}\overline{X}_DX_D)  +\frac{1}{2}g_{\mu\nu}\partial_{\rho}(\overline{X}_Di\gamma^{\rho}X_D) +\frac{1}{2}\nonumber\\
&   \overline{X}_D i(\gamma_{\mu}\partial_{\nu}+\gamma_{\nu}\partial_{\mu}) X_D -\frac{1}{4}\partial_{\mu}(\overline{X}_Di\gamma_{\nu}X_D)
  -\frac{1}{4}\partial_{\nu}(\overline{X}_Di\gamma_{\mu}X_D)\,,\\
 T_{\mu\nu}^{X_V} &= -g_{\mu\nu}(-\frac{1}{4}F_{\rho\sigma} F^{\rho\sigma}
  + \frac{m_X^2}{2} X^{}_{V\rho}X_{V}^{\rho}) +F_{\mu\rho}F^{\rho}_{\nu} +m^2_{X}X^{}_{V\mu}X^{}_{V\nu}\,,
\end{eqnarray}
where $F_{\mu\nu}$ is the field strength tensor. We consider three dark matter spins: a real scalar ($X_R$), a Dirac fermion ($X_D$), and a vector ($X_V$). The interaction Lagrangian with the SM particles is: 
\begin{equation}\label{y-sm}
 {\mathcal L}_{\rm SM}^{Y_2} = -\frac{1}{\Lambda} \sum_i g^{T}_{i}\, T^i_{\mu\nu} Y_2^{\mu\nu}\,,
\end{equation}
where $i$ denotes the SM fields: the Higgs doublet ($H$), quarks ($q$), leptons ($\ell$),
 and $SU(3)_C$, $SU(2)_L$ and $U(1)_Y$ gauge bosons ($g,W,B$). Following~\cite{Ellis:2012jv,Englert:2012xt}, the phenomenological coupling parameters are defined as:
\begin{equation}
 g^T_i=\{g^T_H,\,g^T_q,\,g^T_\ell,\,g^T_g,\,g^T_W,\,g^T_B\}
\label{smcouplings}
\end{equation}
without assuming any UV complete model. Notice that the interaction Lagrangian in Eq.~\eqref{y-sm} defines couplings of the graviton with all SM fields. This hypothesis is more generic with respect to the standard assumptions of the DMsimps, where the mediator interacts only with the quark sector. The energy-momentum tensors of the SM fields are similar to Eqs.~\eqref{em-x} and their explicit expression is provided in \eg~\cite{Das:2016pbk}.  

Complying with the DMsimp idea, it is instructive to consider universal couplings between the spin-2 mediator and the SM particles: 
\begin{align}
 g^T_H=g^T_q=g^T_\ell=g^T_g=g^T_W=g^T_B  \equiv g_{\rm SM}\,.
\end{align}
With this simplification, the model has only four independent parameters\footnote{We have dropped the superscript $T$ for simplicity.},
two masses and two couplings, as for the other DMsimps considered so far:
\begin{equation}\label{param}
 \{m_X,\,m_Y,\,\gdm/\Lambda,\,\gsm/\Lambda\}\,.
\end{equation}
This scenario with a universal coupling to SM particles is realised, \eg, in the original Randall--Sundrum model of localised gravity~\cite{Randall:1999ee}. With this choice of couplings the mediator decays mainly into gluons and light quarks, while the di-photon branching ratio is only $\sim 5$\%. The decay into top-quarks or vector bosons is relevant when kinematically allowed.  As already discussed in the case of spin-1 mediator, the $Y_2$-neutrino coupling leads to $\MET$ signals that are independent of the decays into dark matter particles and provide additional $\MET$ channels for the mono-X signals.

In the following, to exemplify the complementarity of dark matter searches, we will focus on vectorial dark matter. 
\begin{itemize}
\item {\bfseries LHC $\MET$ searches.} The $Y_2$ production is mostly initiated by gluon fusion at low masses, which suppresses mono-photon, mono-Z and mono-W signals, as they can occur only in quark initiated processes. Hence the most constraining missing energy searches for the spin-2 model are a single mono-jet + $\MET$ (ATLAS~\cite{Aaboud:2016tnv}) and 2--6 jets + $\MET$ (ATLAS~\cite{Aaboud:2016zdn}).
\item {\bfseries LHC mediator searches.} Resonance searches from LHC Run 2 data (ATLAS~\citep{ATLAS:2016cyf,ATLAS:2016cwq,ATLAS:2016lvi,ATLAS:2016bvn,ATLAS:2016npe,ATLAS:2016gvq,ATLAS:2016ixk} and CMS~\citep{Khachatryan:2016yec,Sirunyan:2016iap,CMS:2015nmz,CMS:2016abv,Khachatryan:2016qkc,CMS:2016zte}) give strong constraints on the graviton mass in between few hundreds of GeV and several TeV. The considered final states are $jj, ll, \gamma\gamma, W^+W^-, ZZ, hh, b\bar{b}, t\bar{t}$. 
\item {\bfseries Relic density.} The dark matter can achieve the correct relic density via the $s$-channel exchange of a graviton, especially in the region $\mmed \sim 2 \mmed$, and via $t$-channel annihilation into a pairs of $Y_2$, which subsequently decay into SM particles, in the region $\mdm < \mmed$. Both annihilation channels are $s$-wave in the case of vectorial dark matter. The analytic expression for these channels are provided in~\citep{Lee:2014caa}.
\item {\bfseries Indirect detection.} Annihilation via $s$-channel into SM particles with $Y_2$ exchange can produce both a continuum photon spectrum and gamma-ray lines. Both signals can be constrained by Fermi-LAT and HESS spectral feature searches at the Galactic Centre and by Fermi-LAT dSph exclusion limits. Additionally the $t$-channel annihilation process can give rise to box-shaped gamma-ray signatures, see \eg~\citep{Ibarra:2015tya}, which are however only poorly constrained by Fermi-LAT searches for spectral features towards the Galactic Centre~\citep{Lee:2014caa}.
\item {\bfseries Direct detection.} The WIMP-gluon interaction is relevant for direct searches: this coupling generates a twist-2 operator which induces a spin-independent cross-section dark matter-nucleon. This cross section can be in tension with the XENON1T for dark matter masses below roughly 400 GeV, see~\citep{Chu:2012qy,Lee:2013bua} for the case of scalar dark matter. However we couldn't find a dedicated analysis illustrating how direct detection impacts the whole DMsimp spin-2 parameter space. The elastic cross section WIMP-nuclei can receive additional contributions in non minimalistic models~\citep{Lee:2014caa}.
\end{itemize}
The diagrams for dark matter annihilation are illustrated in Fig.~\ref{fig:tabE}, while the mediator production at the LHC is shown in Fig.~\ref{fig:lhcschan}. 
 %
\begin{figure}[t]
\begin{center}
\includegraphics[width=1.\textwidth,trim=0mm 0mm 0mm 0mm, clip]{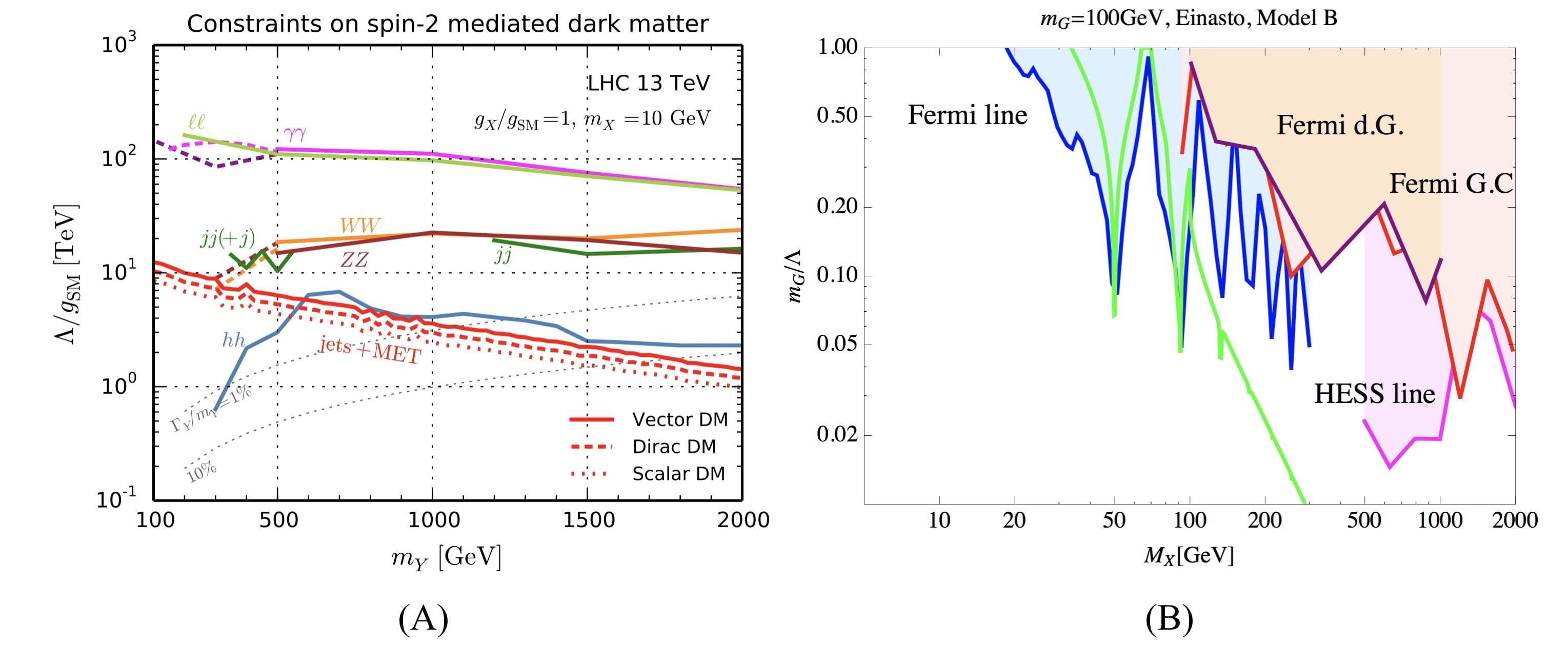}
\end{center}
\caption{{\bfseries DMsimp: $s$-channel spin-2 mediator.} Panel (A): Summary of the 13 TeV LHC constraints in the $\{\Lambda/\gsm, m_Y \}$-plane. The other parameters are fixed as labelled. The differences among the different dark matter spins is not visible in the limits from the resonance searches (as labelled in the plots), conversely to the case of jets + $\MET$ searches (red lines as labelled). Regions below each lines are disfavoured at 95\% CL. Information on the mediator width-to-mass ratio is provided by the grey dotted lines. Figure taken from~\citep{Kraml:2017atm}. The reader can identify $m_Y = \mmed$, $g_X = \gdm$ and $m_X = \mdm$ with respect to the convention used in the review. Panel (B): Gamma-ray bounds from Fermi-LAT (d.G., line, G.C.) and HESS telescope (lines) are shown in case of vector dark matter in the $\{ m_G/\Lambda, M_X \}$-plane, for a fixed graviton mass as labelled. Couplings are not universal, but fixed at $g_X=1, g_V=g_g=g_\gamma=0.3$ and $g_h=0$, and $m_G/\Lambda$ corresponds to the $Y_2$-WIMP coupling $\gdm$. The green line corresponds to the region of parameter space achieving the correct $\relic$. Figure taken from~\citep{Lee:2014caa}. The reader can identify $M_X =\mdm$ and $m_G = \mmed$ with respect to the convention used in the review, while d.G.  and G.C. stand for dSphs and Galactic Center respectively.} \label{fig:spin2}
\end{figure}

At present, to the best of our knowledge, there are actually no comprehensive studies of spin-2 models, which encompass both LHC and dark matter searches, except for~\citep{Zhang:2016xtc}. We however refrain from using their results to illustrate the main features of this model, as they perform a random scan of the full parameter space. While this is certainly instructive, it is not necessarily the most optimal pedagogical approach to begin with. For the sake of the discussion, we choose to show 2D parameter scans, even though they do not show the complementarity of searches.

Figure~\ref{fig:spin2}, from~\citep{Kraml:2017atm,Lee:2014caa}, resumes the constraints on slices of the DMsimp parameter space stemming from LHC searches for a massive graviton, panel (A), and the dark matter gamma-ray searches, panel (B). From panel (A), we clearly see that the di-photon and the di-lepton resonance searches provide the most stringent limit in the whole mediator mass range, constraining $\Lambda/\gsm > 100 \, \TeV$ for graviton masses below 1 TeV. These searches are rather independent on the exact dark matter mass value. Mono-jets + $\MET$ searches become competitive for large values of $\gsm$ and, if the $Y_2$ decays into $\gamma\gamma$ and $ll$, are heavily suppressed. In panel (B), we show the impact of gamma-ray searches. For $\mdm < \mmed$, the exclusion limits from gamma-ray lines provided by Fermi-LAT disfavour at 95\% CL the model parameter space compatible with the thermal relic assumptions, as the dark matter annihilates mainly into $gg$ and $\gamma\gamma$. For $\mdm > \mmed$ the thermal relic scenario is compatible with Fermi-LAT dSph upper limits and with the HESS gamma-ray line searches, which are the most sensitive constraints for large dark matter masses.

Fermionic and scalar dark matter particles are more loosely constrained by current gamma-ray searches with respect to vectorial dark matter particles, as $\sigmav_0$ is suppressed by $p$-wave or $d$-wave respectively. LHC constraints are less sensitive to the dark matter spin. The sensitivity to the dark matter particle nature depends on the hierarchy between $\gdm$ and $\gsm$: for $\gdm \sim \gsm$ only jets + $\MET$ searches can differentiate among the spin of the dark matter candidate; for $\gsm >> \gdm$ all searches become sensitive to the dark matter nature. It turns out that the vectorial case is the most constrained model, while the scalar DMsimp is the less constrained and the fermionic case lies in between.

\subsection{$t$-channel mediator models}\label{sec:tchan}

In this section we discuss the phenomenology of $t$-channel DMsimps and their current state of art with respect to the experimental situation. $t$-channel models couple directly the dark matter sector with the SM fermions (primarily quarks), leading to a different phenomenology with respect to $s$-channel models. The fields in the dark sector are both odd under a $Z_2$ symmetry to ensure the stability of the dark matter candidate, while in $s$-channel models the mediator is usually assumed to be even under the $Z_2$ symmetry.\footnote{In DMsimp $s$-channel models, the mediator cannot be odd under the $Z_2$ otherwise for $\mmed < \mdm$ it would be playing the role of dark matter candidate.} As a consequence, LHC searches are always characterised by $\MET$ signals, as the mediator is produced each time in combination with a dark matter particle. In order to connect the dark matter via $t$-channel with SM quarks there are two main possibilities: scalar dark matter and fermionic mediator, or fermionic dark matter and scalar mediator. The dark matter cannot have colour charge, hence the mediator has to be coloured. Additionally, to comply to MFV, either the mediator or the dark matter should have a flavour index. Here we assume to be the former case. For uncoloured mediator models see~\citep{Garny:2015wea}, while for flavoured dark matter we refer to~\citep{Agrawal:2011ze,Kile:2013ola,Agrawal:2014una}. From the point of view of QCD corrections, the $t$-channel and $s$-channel models are very different, as in the former the mediator can be either neutral or coloured, rendering more involved the treatment of NLO corrections. This has not been yet fully investigated in the literature, due to its complexity.

Among the vast literature on $t$-channel models, see \eg~\citep{Blumlein:1996qp,Cao:2009yy,Barger:2011jg,Bell:2011if,Bell:2012rg,DiFranzo:2013vra,Toma:2013bka,Giacchino:2013bta,Garny:2013ama,An:2013xka,Chang:2013oia,Bai:2013iqa,Bai:2014osa,Chang:2014tea,Garny:2014waa,Ibarra:2014qma,Giacchino:2014moa,Yu:2014pra,Papucci:2014iwa,Giacchino:2015hvk,Bell:2015rdw,Abdallah:2015ter,Abercrombie:2015wmb,Bringmann:2015cpa,Ibarra:2015nca,Carpenter:2016thc,DeSimone:2016fbz,Brennan:2016xjh,Goyal:2016zeh,ElHedri:2017nny,Garny:2018icg}, we choose to present the results obtained in~\citep{Colucci:2018vxz} for the case of scalar dark matter and fermionic mediator, which is the most updated analysis at the time of writing. For the case of fermionic dark matter and scalar mediator we discuss the results presented in~\citep{Garny:2015wea}, which is a comprehensive review paper focusing on $t$-channel simplified models alike the supersymmetric one.

Let us first discuss the case of scalar dark matter candidate $S$ and a vector-like fermionic mediator $T$. We assume that the dark matter is a $SU(2)_L$ singlet, hence it cannot couple at tree level with the weak gauge bosons. Consequently the dark matter hyper-charge is zero, in order to obtain an electrically neutral particle. The dark matter can couple to either right-handed or left-handed SM fermions. Here we assume a couplings with right-handed quarks, in particular only with the third generation. The main reason is dictated by the fact that right-handed couplings to quarks play a major role for the LHC and direct detection phenomenology and the Yukawa of the top is the largest coupling. The mediator $T$ should be a colour triplet, have opposite hyper-charge with respect to the right-handed quarks and be a singlet under $SU(2)_L$. The interaction Lagrangian between WIMPs and the SM quarks is then given by~\citep{Colucci:2018vxz}:
\begin{equation}\label{eq:linttchan}
\mathcal{L}_S^T  = y S \bar{T} P_R t + \rm h.c.\,,
\end{equation} 
where $P_R$ is the right-handed chirality projector, and  we have neglected the quartic term connecting the dark matter particle with the SM Higgs doublet, in the spirit of DMsimp construction. With these assumptions the model has only three free parameters: 
\begin{equation}\label{eq:paramtchan}
 \{\mdm,\, \mmed, \, y\} \,.
\end{equation}
This model considers top-philic dark matter, which might seem more ad hoc than a generic framework where the dark matter couples to all generations. However this is enough to comprehend all the relevant phenomenology, as in the limit $\mdm > m_t$, the results are strictly equivalent as for the case in which the dark matter couples to the light quark (or lepton) generations only. Moreover at energies comparable with the top mass, the computation of QCD and bremsstrahlung corrections are much more involved than in the chiral limit, hence it is relevant to have the most general framework where to treat them. We are not providing any detail on this part and refer to~\citep{Bringmann:2017sko,Colucci:2018vxz} the interested reader.  Notice that if the dark matter was coupled to all three quark generations with three different vector-like fermionic mediators, MFV requirements would enforce the three mediator masses to be equal, as well as their couplings with WIMPs and quarks.

\begin{figure}[t]
\begin{center}
\includegraphics[width=0.7\textwidth,trim=0mm 0mm 0mm 0mm, clip]{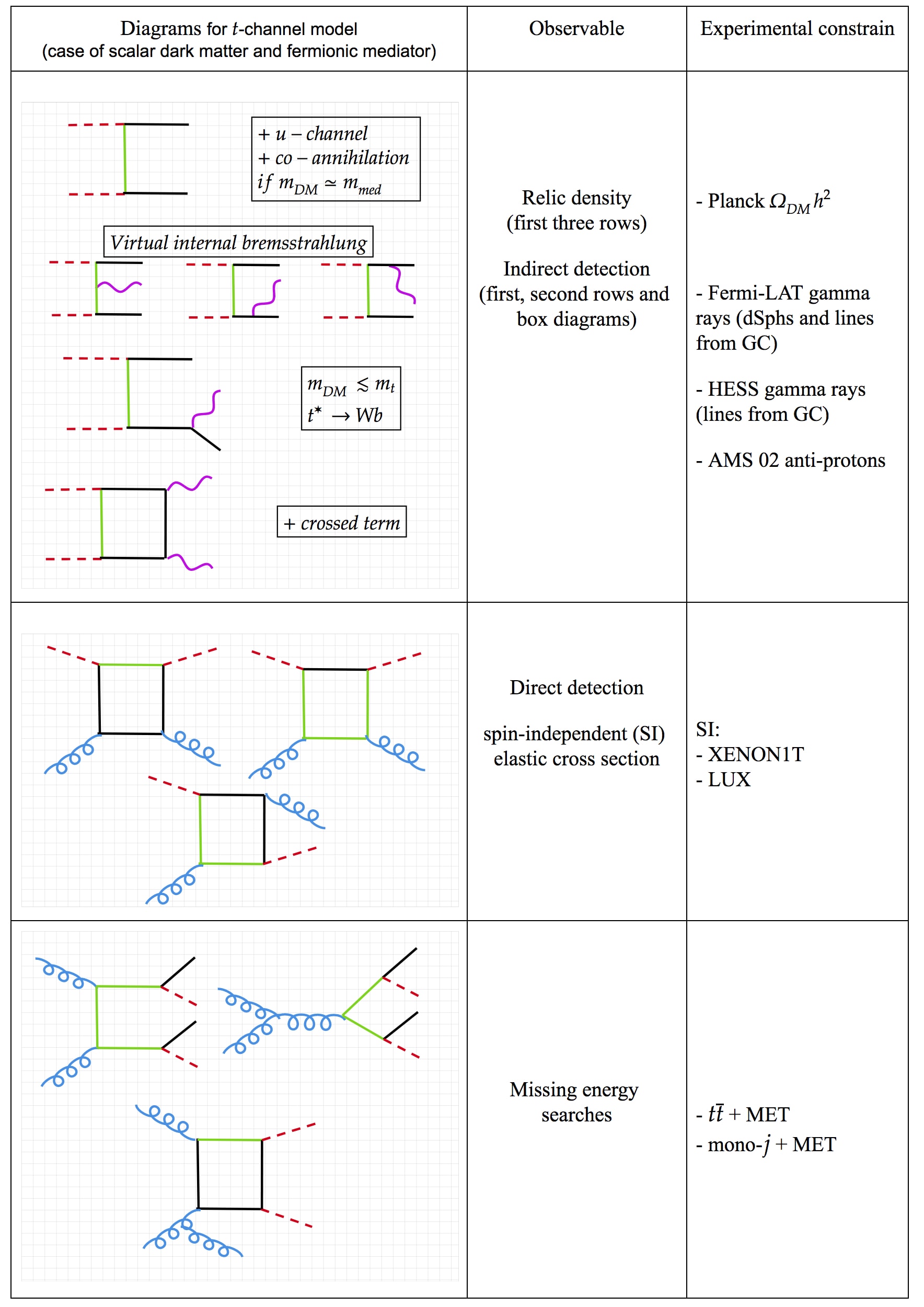}
\end{center}
\caption{Schematic of leading order diagrams contributing to all dark matter searches in the $t$-channel DMsimps with scalar dark matter and fermionic mediator. The diagrams contributing to LHC searches are specifically drawn for the case of the top-philic model discussed in Sec.~\ref{sec:tchan} (for generic fermionic mediator the reader is referred to~\citep{DeSimone:2016fbz} and the references therein). The case of Majorana dark matter and scalar mediator is easily obtained from the above diagrams. For fermionic dark matter there is an additional spin-dependent contribution to the direct detection elastic scattering cross section. MET stands for missing transverse energy. The colour code is as in Fig.~\ref{fig:diagschans0s1}.}\label{fig:tchan}
\end{figure}

The dark matter constraints for this model are:
\begin{itemize}
\item {\bfseries LHC searches with $\MET$.} There are two types of searches particularly relevant for this model: (i) supersymmetric searches of scalar top partners (LEP~\citep{Abbiendi:2002mp} and LHC~\citep{Sirunyan:2017leh,CMS:2017odo}), recasted to constrain the vector-like fermionic mediator of the model, which is strongly interacting and leads to mainly $t\bar{t} + \MET$ signals; (ii) the usual dark matter searches characterised by a mono-jet + $\MET$~\citep{Aaboud:2016tnv,Aaboud:2017phn,Sirunyan:2017hci,Sirunyan:2017jix} (actually the most updated mono-$j$ searches do include more than one hard jet). NLO QCD corrections and matching with the parton showers have been taken into account, in order to comply with the state-of-art modelling of the LHC signals for the $s$-channel case. 
\item {\bfseries Relic density.} There are several annihilation processes contributing to $\relic$, depending on the model parameter space region. For $\mdm >> m_t$ the chiral limit is valid and virtual internal bremsstrahlung (VIB) adds a significative contribution to the tree level leading order $t$-channel diagram, which is helicity suppressed and the first non-zero term depends on $v^4$ ($d$-wave). Decreasing $\mdm$ just above the top threshold, the tree level $t$-channel diagram, which is $s$-wave, is the leading contribution to $\sigmav_0$. Below the top mass, loop-induced processes into $\gamma\gamma$ and $gg$ can play a role (similarly to the spin-0 top-philic dark matter presented in Sec.~\ref{sec:spin0}), while, for $\mdm \lesssim m_t$ the off-shell decay $t^\ast \to W b$ is relevant. Additionally, if the dark matter and the mediator masses are close in mass (within 10\%) co-annihilation between $S$ and $T$ is also relevant, as well as $T$ annihilations. 
\item{\bfseries Indirect detection.} For $\mdm < m_t$, annihilation via the loop induced process into pairs of gluons dominates. This leads to a prompt photon spectrum. The $\gamma\gamma$ final state is subdominant with respect to the $gg$ final state as already discussed in Sec.~\ref{sec:spin0}, however it gives rise to box-shaped gamma-ray signals (the width of the box depends on the mass hierarchy between $S$ and $T$: if they are quasi degenerate the box is very narrow, otherwise it is a wide box). For $\mdm \sim m_t$, the dominant annihilation channel is the tree level $t$-channel, $SS \to t\bar{t}$, which leads to a continuum spectrum of prompt photons, detectable by the Fermi-LAT dSph searches. The same process can be constrained with the anti-proton data released by AMS 02. VIB with the emission of a photon,  a gluon or a weak boson, has been demonstrated to be the dominant contribution in the chiral limit ($\mdm >> m_t$), see \eg~\citep{Giacchino:2013bta,Giacchino:2014moa,Giacchino:2015hvk,Bringmann:2015cpa,Bell:2011if,Bringmann:2017sko}. The emission of an additional vector boson lifts the helicity suppression and gives rise to sizeable $\sigmav_0$. If $S$ and $T$ are nearly degenerate in mass, the $SS \to t\bar{t} \gamma$ process dominates the VIB contribution. This photon emission gives rise to a sharp spectral feature, that can be constrained with current gamma-ray line searches. Indeed, the present telescope resolution does not allow to discriminate among the sharp edge due to VIB or a true gamma-ray line~\citep{Garny:2015wea}. Direct annihilation of the dark matter into photon pairs via box diagram is on the same foot as VIB. On the other hand the annihilation process $S S \to t \bar{t} g$ contributes to the continuum photon spectrum.
\item {\bfseries Direct detection.} An effective coupling WIMP-gluons generates a spin-independent contribution to the elastic scattering cross section, which is, for $\mdm < m_t$, in tension with the XENON1T bound. Conversely for $\mdm > m_t$, $\sigma^{\rm SI}_n$ is negligible and below the neutrino background~\citep{Billard:2013qya}.
\end{itemize}
The relevant diagrams contributing to all dark matter searches in this DMsimp are shown in Fig~\ref{fig:tchan}, while the dependency on the dark matter spin is summarised in Tab.~\ref{tab:Dmnat}. 

\begin{figure}[t]
\begin{center}
\includegraphics[width=1.\textwidth,trim=0mm 0mm 0mm 0mm, clip]{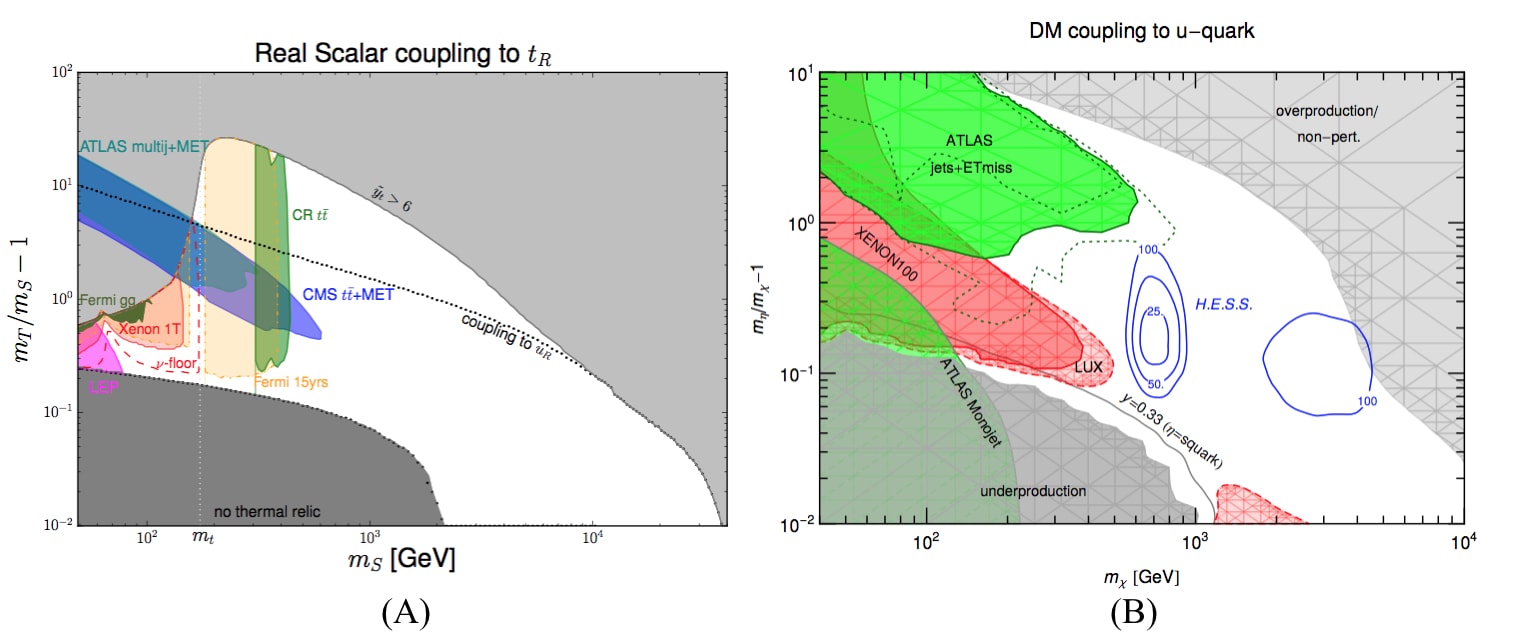}
\end{center}
\caption{{\bfseries DMsimp: $t$-channel.}  Panel (A): Combined constraints from direct and indirect detection and collider searches in the $\{ (m_T/m_S-1), m_S\}$-plane. The grey regions stand for either under-abundant or for over-abundant dark matter for fermionic mediator and scalar WIMPs. The red region is excluded at 90\% CL by XENON1T, while the red dashed line indicates the region of parameter above the neutrino floor hence detectable by direct detection. The green regions are excluded by Fermi-LAT dSph constraints on gamma rays and by anti-protons~\citep{Cuoco:2017iax}. The orange region denotes the expected sensitivity of 15 years of data taking by Fermi-LAT. The magenta and blue regions show constraints on scalar top production at LEP and at the LHC, while mono-X + $\MET$ searches disfavour the dark blue region at 95\% CL. Figure taken from~\citep{Colucci:2018vxz}. The reader can identify $m_T = \mmed$ and $m_S = \mdm$ with respect to the convention used in the review. Panel (B):  Same as panel (A) for Majorana dark matter and scalar mediator. The colour code is: grey regions denote under- and over-abundant dark matter, while the green regions are disfavoured at 95\% CL by jet(s) searches + $\MET$. The red regions are excluded by XENON100 and LUX at 90\% CL. The blue contour levels show the ratio between the excluded annihilation cross section and the thermal cross section. The regions inside these contour lines are disfavoured if the gamma-ray flux from dark matter annihilation is enhanced relative to the Einasto profile~\citep{1989A&A...223...89E} by the corresponding factor (more cuspy profiles or presence of substructures). Figure taken from~\citep{Garny:2015wea}. The reader can identify $m_\chi =\mdm$ and $m_\eta = \mmed$ with respect to the convention used in the review.} \label{fig:tchannel}
\end{figure}

The results of the comprehensive dark matter study are illustrated in panel (A) of Fig.~\ref{fig:tchannel}. Under the assumption that the dark matter is a thermal relic, the complementarity of dark matter searches is clearly shown in the plot. Direct detection experiments probe the region for $\mdm < m_t$, while Fermi-LAT, HESS and AMS 02 are sensitive to a mass range from roughly $m_t$ up to 500 GeV. This shows that anti-matter constraints can be competitive with gamma-ray searches, modulo the larger astrophysical uncertainties. LEP searches constrain the most lightest values of $\mdm$, while CMS searches cover a parameter space orthogonal  to indirect detection. In particular multi-jets + $\MET$ searches loose quickly sensitivity with the increase of the dark matter mass, however the $t\bar{t} + \MET$ searches are effective in the regime where the decay $T \to S t$ happens far from threshold. Notice that if the decay channel $T \to S t$  is closed, the mediator becomes long-lived. This case requires further dedicated studies. 

The Majorana dark matter DMsimp exhibits only few differences with respect to the scalar dark matter model presented above. We summarise here the most important. Under the same assumptions made for the fermionic mediator, the interaction Lagrangian with only a single generation of light quark (considering the model in~\citep{Garny:2015wea}) is given by:
\begin{equation}
\mathcal{L}_S^T  = y \tilde{T}^{\ast} \bar{X} P_R q + \rm h.c.\,,
\end{equation}
where now the dark matter field is denoted by $X$ and the mediator by $\tilde{T}$ and $q$ is the light quark, which we assume to be the $u$ flavour for concreteness for the rest of the section. This Majorana model is very close to the simplified model considered in supersymmetric searches at the LHC, as it is implemented in the Minimal Supersymmetric Model with only light quarks and the neutralino, except that the coupling $y$ is not fixed at the weak scale but can be varied freely.

\begin{itemize}
\item {\bfseries LHC searches with $\MET$.} As for the fermionic dark matter case, the most stringent searches are given by jets + $\MET$ and mono-jet + $\MET$. The first arises from the direct production of the coloured mediator that further decays into the dark matter and light quarks. The latter stems from the loop-induced production of a dark matter pair that recoils against a jet.
\item {\bfseries Relic density.} The annihilation processes contributing to the dark matter relic density are analogous to the case of scalar dark matter in the chiral limit. In this limit, the $X X \to u_R u_R$ process is $p$-wave suppressed.
\item{\bfseries Indirect detection.} This is completely analogous to the scalar dark matter case in the chiral limit, except that the $t$-channel tree level annihilation diagram is still helicity suppressed and the first non zero term in the chiral limit is $p$-wave. The authors in~\citep{Garny:2015wea} consider as well the exclusion limits on $\sigma^{\rm SD}_p$ stemming from IceCube.  The vector boson generating the neutrino flux from the Sun arises from VIB: $X X \to u_R u_R V$ and subsequently $V=W,Z,h$ shower and hadronise and produce a continuum spectrum for the neutrinos. The IceCube bounds are less performant that direct detection searches to constrain the model parameter space, hence are not shown in the following.
\item {\bfseries Direct detection.} With respect to the diagrams shown in Fig.~\ref{fig:tchan}, the elastic cross section off nucleus receives an additional contribution from the $s$-channel exchange of $\tilde{T}$, which is not present in the scalar dark matter case. There is a small contribution from the spin-independent operator, while the  leading contributions to the elastic cross section are proportional to a combination of $\mathcal{O}_4^{\rm NR}$, $\mathcal{O}_8^{\rm NR}$ and $\mathcal{O}_9^{\rm NR}$, which are spin-dependent operators. Still the most constraining bounds on the model come from spin-independent limits from the LUX experiment, as these are orders of magnitude more sensitive than the spin-dependent upper bounds.
\end{itemize}

The results of the comprehensive dark matter study are illustrated in panel (B) of Fig.~\ref{fig:tchannel}. The picture is rather similar to the case of scalar dark matter, assuming a thermal dark matter scenario. Constraints from jets + $\MET$ are relevant for large mass splitting between $X$ and $\tilde{T}$, because after its production the mediator has a larger phase space for its decay into the dark matter and the light quark, leading to harder jets. Direct detection is sensitive to smaller mass splitting, while mono-jet +$\MET$ searches are sensitive to the quasi degenerate region. Gamma rays probe the model parameter space in the intermediate $\mdm$ mass range.
%
%

The Dirac dark matter DMsimp is different from the Majorana case reported above, as far as it concerns the dark matter studies. The elastic scattering cross section is dominated by spin-independent because of the contribution from vectorial currents, which are null in case of Majorana fermions. Hence, thermal Dirac dark matter models get strongly constrained by current direct detection experiments, which combined with LHC searches, completely disfavour the thermal hypothesis for such model, see~\citep{DeSimone:2016fbz} for details.

As a concluding remark, in general a coloured $t$-channel mediator scenario will be probed to a large extent by next generation experiments, assuming thermal dark matter production and perturbativity of the coupling.

\section{Caveats of dark matter simplified models}\label{sec:caveats}

DMsimps represent an improvement with respect to the use of EFT for collider dark matter searches during LHC Run 1. However most of them are still considered not the ideal benchmark models over which categorise the dark matter searches and their complementarity. The main reason of concern is related to the fact that most of the DMsimps are not gauge invariant, thus not renormalisable, see \eg~\citep{Kahlhoefer:2015bea,Bell:2015sza,Bell:2015rdw,Bell:2016uhg,Englert:2016joy,Haisch:2016usn,Bell:2016fqf}. The most striking example is provided by the spin-1 mediator with axial-vector couplings to fermions. The interaction Lagrangians provided in Sec.~\ref{sec:spin1} are not gauge invariant unless a dark Higgs is introduced to give mass to the $Z'$ mediator, and hence to unitarise its longitudinal component. As a consequence, the most minimalistic self consistent model would feature two mediators, an additional scalar along with the $Z'$. The presence of a second mediator would change the phenomenology of the model, which is not anymore well described by the single mediator assumption. 

The $s$-channel scalar mediator case is not gauge invariant unless $Y_0$ mixes with the Higgs boson, because the dark matter is a singlet under the SM gauge symmetries, see \eg~\citep{Lopez-Val:2014jva,Khoze:2015sra,Baek:2015lna,Wang:2015cda,Costa:2015llh,Robens:2016xkb,Dupuis:2016fda,Balazs:2016tbi}. The mixing with the Higgs boson introduces a major modification in the building of the next generation of DMsimps, as the model parameter space then becomes constrained by measurements of the Higgs properties. This has motivated two types of scenarios: (i) models that communicate with the SM via the Higgs portal through the mixing parameters, or even models for which the scalar mediator is the Higgs itself; (ii) to avoid the tight constraints stemming from Higgs physics, $Y_0$ mixes with an additional doublet similarly to a two Higgs doublet model. Likewise, pseudo-scalar DMsimps~\citep{Goncalves:2016iyg} can be made theoretically consistent by promoting them to double mediator models. Two Higgs doublet models are well motivated theoretically, arising in several UV complete models such as supersymmetry, or other extensions of the SM, see \eg~\citep{Fayet:1976et,Gunion:1984yn,Amaldi:1991cn,Carena:1995wu,Branco:2011iw,Bhattacharyya:2015nca} and the references therein.

We will not discuss more in details here these issues and the proposed solutions. There is already a quite vast literature along the lines of the two Higgs doublet models and Higgs portals. The interested reader is referred to \eg~\citep{Bauer:2016gys,DeSimone:2016fbz,Duerr:2016tmh,Boveia:2016mrp,Ko:2016zxg,Bell:2016ekl,Bell:2017rgi,Bauer:2017ota,Ellis:2017tkh,Baek:2017vzd,Sanderson:2018lmj}.

\section{Future prospects}\label{sec:fut}

The focus of this review has been to describe the state-of-art of dark matter simplified models, as defined by the LHC Dark Matter Working group, with respect to the current dark matter searches. In particular we have discussed the degree of complementarity of LHC searches (mono-jet + $\MET$, jets + $\MET$, resonance searches), dark matter direct and indirect detection searches (gamma-ray, anti-matter and neutrino searches) in several scenarios: $s$-channel mediator with spin-0 and spin-1 and Dirac dark matter, $s$-channel mediator with spin-2 and vectorial dark matter, $t$-channel mediator with either scalar dark matter and fermionic mediator or vice-versa.

DMsimps provide a simple framework where to define, categorise and compare dark matter searches. These comprehensive analyses are a powerful tool to understand the dynamic underlying the various dark matter searches, modulo their interpretation being  subject to the caveats described in the previous section. Keeping in mind the main assumption that the dark matter and the mediator are the only particles of the dark sector accessible at current experiments, we can formulate few general statements from the global analyses presented in this review.
\begin{itemize}
\item LHC searches for dark matter with mono-X and missing energy, direct dark matter searches and indirect dark matter searches in general probe regions of the model parameter space which are complementary to each other. Typically indirect searches extend to heavier dark matter masses with respect to LHC and direct detection. Direct detection has better sensitivity than LHC searches in the intermediate dark matter mass range, while LHC performs better in the small dark matter range, where however the dark matter is often not viable as thermal relic (it is either over-abundant or under-abundant, depending on the model). The relic density constraint can be avoided by assuming for instance a non thermal dark matter scenario. Because of the complementarity of searches, for instance a non-detection in direct detection does not preclude a positive detection at the LHC or at gamma-ray telescopes.
\item Some models are already disfavoured as thermal relic by the combination of LHC and direct detection searches (see $t$-channel model with Dirac dark matter or $s$-channel spin-1 with Majorana dark matter). By reversing the argument, we can assert that if a signal is seen in the mono-X + $\MET$ searches, the thermal dark matter hypothesis is under test. This can be solved: (i) by invoking a more complex dark sector, where co-annihilation and new annihilation channels can open up the thermal relic parameter space; (ii) dark matter is produced via additional non-thermal mechanisms to dilute/increase its relic abundance down/up to the observed value.
\item In case of a positive signal at the LHC in a SM + $\MET$ channel, the identification of the dark matter is non-trivial, conversely to the characterisation of the mediator. Luckily all dark matter searches, even though they feature a certain degree of complementarity, also probe common regions of the parameter space. In an optimistic scenario, a signal can be detected in multiple experiments allowing to pin point both the nature of the dark matter and the characteristics of the model. 
\item The spin-2 mediator model has been poorly investigated so far and deserves future careful comprehensive analyses.
\end{itemize}

It is crucial to keep continuing looking for dark matter with a comprehensive approach relying in simplified bottom-up scenarios. Some theoretical shrewdnesses are in place. The use of gauge invariant models certainly constitutes a must, however theoretical predictions can be improved along other directions, which are often neglected. For instance, the wide separation of scales involved in constraining WIMP models, from the LHC to indirect detection and to direct detection, is often neglected. The authors in~\citep{DEramo:2014nmf} have shown that the running of EFT operators from  the  mediator  mass  scale  to  the  nuclear  scales  probed  by direct searches via one-loop Renormalisation Group Equations (RGEs) has an impact for models that would in generally not be constrained by direct detection searches because suppressed by the momentum transfer or by the WIMP velocity.  These models can be excluded as a consequence of spin-independent couplings induced by SM loops. 

Experimentally, the close future is quite promising as there is a rich program expected to start soon and produce results in the next decade or so. Concerning the future of direct detection, starting from 2019, there are several experiments planned able to probe WIMP-nucleon cross section of the order of the neutrino floor ($\sigma^{\rm SI}_n \sim 10^{-48} \rm cm^2$ for $\mdm \sim 30 \GeV$), see XENONnT~\citep{Aprile:2015uzo}, LZ~\citep{Mount:2017qzi} and DARWIN~\citep{Aalbers:2016jon}. At low WIMP mass, around 3-4 GeV, exciting progresses are expected by SuperCDMS SNOLAB~\citep{Agnese:2016cpb}, by CRESST III~\citep{1742-6596-718-4-042048} and by EDELWEISS-III~\citep{Arnaud:2017usi}, which can probe elastic spin-independent cross sections as low as $10^{-44} \rm cm^2$. Concerning indirect detection, the Cherenkov Telescope Array (CTA)~\citep{Acharya:2013sxa} is one of the major advancements in the gamma-ray searches  as it will be sensitive to the energy range in between 20 GeV to 300 TeV. Starting from 2022, while operating, it will provide unprecedented complementary results to direct detection and LHC searches, as it will be sensitive to dark matter masses up to 100 TeV. These future probes, together with the LHC Run 3 foreseen for 2021, can vastly extend the coverage of the dark matter parameter space of simplified models, see~\eg Refs.~\citep{Baum:2017kfa,Balazs:2017hxh,Bertone:2017adx}. The $t$-channel model parameter space, under the hypothesis of thermal dark matter, will be for instance almost entirely probed. 

Lastly, the Sun has been recently proposed as target to constrain a specific class of DMsimps, in which the mediator is light  (MeV range) and long-lived~\citep{Arina:2017sng,Leane:2017vag}. LHC searches are insensitive to this type of mediators, which can however be observed in gamma rays. The Sun is opaque to all dark matter annihilation products but neutrinos and the neutral and weakly interacting mediators (the mechanism that produces the mediators inside the Sun is the same as for the neutrino signal). If these mediators are long-lived enough to decay outside the Sun, they could lead to characteristic gamma-ray signatures detectable within 10 years of Fermi-LAT mission, and in one year of full exposure of ground water Cherenkov telescopes (HAWC~\citep{Abeysekara:2013tza} and LHAASO~\citep{Zhen:2014zpa,He:2016del}). Models with long-lived MeV mediators are actually very constrained by beam dump experiments and cosmology~\citep{Arina:2017sng}. Their entire parameter space can be probed by next generation of intensity experiments, such as NA62~\citep{Dobrich:2017yoq} and SHiP~\citep{Alekhin:2015byh}. Hence dark matter simplified models not only serve as benchmark for high-energy studies but they can be exploited as a bridge relying the high-energy frontiers with the intensity frontiers.

\section*{Author Contributions}

The author confirms being the sole contributor of this work and approved it for publication.

\section*{Funding}

This review article has been supported by the Innoviris grant ATTRACT Brains for Brussels 2015 (BECAP 2015-BB2B-4).

\section*{Acknowledgments}
The author would like to acknowledge Fabio Maltoni, Luca Mantani and Kentarou Mawatari for useful discussions about various aspects of the dark matter simplified models. She is also grateful to Jan Heisig for a careful reading of the manuscript and for providing useful comments.

\bibliographystyle{frontiersinHLTH&FPHY} 
\bibliography{biblio}

\end{document}